\documentstyle[12pt,epsfig]{article}
\font\tenrm=cmr10

\font\elevenbf=cmbx10 scaled\magstep 1
\font\elevenrm=cmr10 scaled\magstep 1
\newcommand{\be}{\begin{equation}}
\newcommand{\ee}{\end{equation}}
\newcommand{\bea}{\begin{eqnarray}}
\newcommand{\eea}{\end{eqnarray}}

\renewenvironment{thebibliography}[1]
{\tenrm
\begin{list}{\arabic{enumi}.}
{\usecounter{enumi}     \setlength{\parsep}{0pt}
\setlength{\itemsep}{3pt} \settowidth{\labelwidth}{#1.}
\sloppy
}}{\end{list}}

\begin{document}

\begin{center}
{\large \bf 
PENTAQUARKS IN CHIRAL SOLITON MODELS
\footnote{Based partly on the talks presented at the International Seminar on High Energy Physics 
Quarks-2004, Pushkinogorie, Russia, May 24-30, 2004; International Workshop on Quantum Field Theory 
and High Energy Physics QFTHEP-04, Saint-Petersburg, Russia, June 17-23, 2004, and 
Symposium of London Mathematical Society "Topological Solitons and their Applications",
Durham, UK, August 2-12, 2004. Slightly reduced version of this paper is available as
E-print HEP-PH/0507028.}.\\}
\vspace{4mm}
{\elevenrm Vladimir B. Kopeliovich\\ 
Institute for Nuclear Research of Russian Academy of Sciences,\\
 Moscow, 117312}
\end{center}
\begin{abstract}
\tenrm\baselineskip=11pt
The spectra of pentaquarks, some of them being observed recently, are discussed within 
topological soliton
 model and compared with simplified quark picture. 
Results obtained within chiral soliton model depend to some extent on the quantization 
scheme: rigid rotator, soft rotator, or bound state model.
The similarity of spectra of baryon resonances obtained within quark model
and chiral soliton model is pointed out, although certain differences take place as well, which 
require careful interpretation. In particular, considerable variation of the strange antiquark mass 
in different $SU(3)$ multiplets of pentaquarks is required to fit their spectra obtained from
chiral solitons. Certain difference of masses of "good" and "bad" diquarks is required as well,
in qualitative agreement with previously made estimates. 
The partners of exotic states with different values of spin
which belong to higher $SU(3)$ multiplets, have the energy considerably higher than states with 
lowest spin, and this could be a point where the difference from simple quark models is striking. 
The antiflavor excitation energies for multibaryons are estimated as well, and binding energies
of $\Theta$-hypernuclei and anticharm (antibeauty) hypernuclei are presented for several baryon 
numbers. Some deficiencies in the argumentation against validity of the chiral soliton approach
and/or $SU(3)$ quantization models, existing in the literature, are pointed out. 
\noindent
\end{abstract}
%
\baselineskip=13pt
\section{Introduction}
Recent events around possible discovery of pentaquarks and negative results, obtained lately,
make situation with the observation of these new type baryonic resonances quite dramatic. 
Indeed, one of striking events in elementary particle physics of the last few years was
observation of baryon resonances with unusual properties (end of 2002 - 2005):

$\Theta^+(1540),\;$  strangeness $S=+1$, isospin $I=0$ (most likely), width 
$\Gamma_\Theta < 10\,MeV$,
seen by different collaborations in Japan, Russia, USA, FRG, CERN;

$\Phi/\Xi_{3/2}^{--}(1862)$, strangeness $\;S=-2\;$, $\;I=3/2\;(?)$, $\Gamma < 18 \,MeV$ observed 
by NA49 Collab.
at CERN\footnote{In the latest issues of PDG the state $\Xi_{3/2}$ is called
 $\Phi$ \cite{PDG}, although the notation $\Xi_{3/2}$ seems to be more informative; here it will be 
denoted as $\Phi/\Xi_{3/2}$ };

$\Theta^0_c(3099)$, charm  $\;C=-1$, $\Gamma < 15\,MeV$ seen by H1 Collab., DESY.
Spin-parity $J^P$ of these states is not measured yet.

These states are manifestly exotic because they cannot be made of 3 valence quarks only.
There are different possibilities to have exotic baryon states:
\\
a) positive strangeness $S>0$ (or negative charm $C<0$, or positive beauty), since $s$-quark has 
$S=-1$ and $c$-quark $C=+1$,
\\
b) large (in modulus) negative strangeness  $S<-3B$, $\;B$ - baryon number; similar for charm 
or beauty, 
\\
c) large enough isospin $I> (3B+S)/2\,$,  if $\;\; -3B \leq S \leq 0$, or
charge $Q\;>\;2B+S$ or $Q\;<\, -B$ in view of Gell-Mann - Nishijima relation $Q=I_3+Y/2$.

The pentaquarks $\Theta^+(1540)$ and $\Theta_c(3099)$, if it is confirmed, are just of the type 
{\bf a)},
the possibility {\bf b)} is difficult to be realized in practice, since large negative strangeness of
produced baryon should be balanced by corresponding amount of positively strange kaons; high enough 
energy of colliding particles is necessary here.
The state $\Phi/\Xi^{--}_{3/2}$ is of the type {\bf c)}.

The minimal quark contents of these states are:

$\Theta^+=(dduu\bar{s});\;  \; \Xi^{--}=(ssdd\bar{u});\;
\Theta_c^0=(dduu\bar{c})$, and by this reason they are called pentaquarks.

The history and chronology of pentaquarks predictions and discovery
 has been discussed already in 
many papers, here I recall it briefly for completeness. Readers familiar with this subject can
go immediately to sections 3, 4.

$\Theta^+,\;S=+1$ baryon was observed first at 
SPring-8 installation  (RCNP, Japan)\cite{1} 
in reaction
$\gamma ^{12}C\, \to K^+n+...$. The reported mass is $1540\pm 10\;MeV$ and width $ \Gamma < 25\,MeV$, 
confidence level (CL) $4.6\sigma $. Soon after this and independently 
DIANA collaboration at ITEP, 
Moscow \cite{2} reported on observation of $\Theta^+$ state in 
interactions of
 $K^+$ in $Xe$ bubble chamber. The mass of 
the bump in $K^0p$ invariant mass distribution is $1539\pm 2\;MeV, \; \Gamma < 9 \,MeV$, 
confidence level a bit lower, about $4.4\sigma$.

Confirmation of this result came also from several other experiments 
\cite{3,4,5,6,7,8,9,10,troyan}
mostly in reactions of photo-(electro)-production.
 The reported mass of $\Theta^+$ is smaller
by several $MeV$, or even by $\sim (10-20)\,MeV$ in some of experiments \cite{5,7,8,9,10}, than first 
reported value $1540\,MeV$ \cite{1}.
The CLAS Collaboration \cite{bdv} provided recently evidence for
two states in $\Theta$-region of the $K^+n$ invariant mass distribution at $1523\pm 5\,MeV$ 
and $1573\pm 5\,Mev$. The nonobservation of $\Theta^+$ in old kaon-nucleon scattering data provided 
restriction on the
width of this state.
 Phase shift analysis of $KN$-scattering in the energy interval  
$1520\,-\,1560\,MeV$ gave a restriction $\Gamma < 1\,MeV$ \cite{asw}. 
Later analysis of data
\cite{2}, obtained in $Xe$ bubble chamber, allowed to get the estimate for
the width of $\Theta^+$:
  $\Gamma_\Theta \simeq 0.9\pm 0.3\,MeV$, and from total cross section 
data $\Gamma_\Theta$ in the interval  $1\,-\,4\,MeV$ \cite{gib}.

Several experiments, mostly at higher energies, did not confirm existence of $\Theta$, pessimistic 
point of view was formulated, e.g., in \cite{poc}. 
More complete list of references to negative 
results, many of them being not published yet,  and their critical discussion can be found in 
\cite{hicks,azim,hic2}.
 Most serious seems to be
recent negative result on $\Theta^+$ photoproduction on protons, obtained at spectrometer CLAS
(JLAB) with high statistics, where no resonance has been observed in $K^+n$ final state in
the mass interval $1520-1600\,MeV$ scanned with $5\,MeV$ steps \cite{devita}.

The doubly strange cascade hyperon $\Phi/\Xi_{3/2},\; S=-2$, probably with isospin $I=3/2$, is observed 
in one experiment at CERN, only, in proton-proton collisions at $17 \,GeV $\cite{alt}. The mass of 
resonance in $\Xi^-\pi^-$ and $\bar{\Xi}^+\pi^+$ systems is $1862\pm 2\;MeV$ and mass of resonance in 
$\Xi^-\pi^+,\;\bar{\Xi}^+\pi^-$ systems is
$1864\pm 5\;MeV,$ width $ \Gamma < 18\,MeV$, and $CL=4.0 \sigma$.
The fact makes this result more reliable, that resonance $\Phi/\Xi^{--}$ was observed in antibaryon 
channel as well.
However, this resonance is not confirmed by HERA-B, ZEUS, CDF, WA-89, COMPASS collaborations (see e.g. 
\cite{poc,comp}), although there is
no direct contradiction with NA49 experiment because other reactions have been used and mostly at 
higher energies, so, upper bounds on the production cross sections of
$\Phi/\Xi_{3/2}$ have been obtained in this way, see \cite{comp,spen} for compilation of these results.

The anticharmed pentaquark $\Theta_c^0,\;\; C=-1$ was observed at
 H1, HERA, Germany, in
$ep$ collisions \cite{akt}, in both baryon number $B=\pm 1$ channels. The mass of resonance in 
$D^{*-}p,\;D^{*+}\bar{p}$ systems is $3099\pm 3\pm 5\;MeV,
 \; 
\Gamma < 15\,MeV, \; CL= 6.2\sigma$.
ZEUS collaboration at HERA \cite{zeus} did not confirm the existence of $\Theta_c^0$ with this 
value of mass, and
 this seems to be serious contradiction, see again \cite{poc}.

There is also some evidence for existence of other baryon states, probably exotic, e.g. nonstrange 
state decaying into nucleon and two pions \cite{rip}, and resonance in  $\Lambda K_S^0$ system with 
mass $1734 \pm 6\,MeV,\;
  \Gamma \,<\, 6\,MeV$ at $CL=(3 - 6)\sigma$
which is $N^{*0}$ or $\Xi^*_{1/2}$,
 observed by STAR collaboration at RHIC \cite{kab} in reaction
$Au+Au$ at $\sqrt{s_{NN}}\sim 200\,GeV$.
 Several resonances in $\Lambda K_S^0$ system have been 
observed recently at JINR \cite{aer}, the lowest
one has the mass $1750 \pm 18\,MeV$, not in contradiction with \cite{kab}.

Evidently, high statistics experiments on pentaquarks production are drastically needed, as 
well as checking the relatively old scattering data analyzed in \cite{asw,gib}. If high statistics 
experiments do not confirm existence of $\Theta^+$, it 
would be interesting then to understand 
why more than 10 different experiments, although each of them with 
not high 
statistics, using different installations and incident particles, provided 
similar
 positive results.
 From theoretical point of view the interest to such exotic baryon states 
will not dissappear in any case, because they represent the next in complexity step after baryons
made of 3 valence quarks and should most probably appear at higher masses and with greater widths.

 Information about status of higher statistics experiments performed or to
be performed at JLAB (CLAS Collab.) can be found in \cite{pr,hic2}. Several reviews of existing 
experimental situation appeared lately, e.g. \cite{comp,spen,azim,hic2,kab2}, and I will not 
go into further details here.
In any case, the difficulties in observation of such exotic states mean that the role of these
states in hadron dynamics of moderate and especially high energies is not big and cannot be even
compared with, e.g., the role of $\Delta(1232)$ resonance in medium energy pion-nucleon interactions.

Next sections necessarily contain certain overlap with previous discussions, I hope to add some 
new accent of criticism to the discussion of this interesting topic. Sections 3,4 contain short 
description of the topological soliton models, in section 5 the results for baryon spectra are
presented.
In Section 6 the large $N_c$ arguments that the width of $\Theta^+$ is expected to be greater than width
of $\Delta$ resonance are criticized and ambiguity of large $N_c$ consideration is stressed, in 
Section 7 a correspondence with the quark model description is established and some 
difference is fixed and discussed as well. In section 8 the masses of partners of lowest pentaquark 
states with different spin or isospin are estimated.
In Section 9 the multibaryons with additional quark-antiquark pairs are
discussed within chiral soliton approach, which can appear as $\Theta$ hypernuclei, or hypernuclei
with anti-charm (-beauty), and it is argued that existence of such states is a 
natural property of this approach. Section 10 contains conclusions and some prospects.
\section{Early predictions}
From theoretical point of view the existence of such exotic states by itself was not unexpected.
Such states have been discussed first by R.L.Jaffe within MIT quark-bag model
 \cite{j}. 
The mass of these states 
was estimated to be considerably higher than that reported now:
$M_\Theta \simeq 1700 \;MeV, \;\; J^P=1/2^-$.
These studies were continued by other authors \cite{hog}:
$M_\Theta \simeq 1900\,MeV$, $J^P=1/2^-$, and similar in \cite{str}.
From analysis of existed that time data on $KN$ interactions the estimate was obtained in \cite{roi}
$M_\Theta \simeq 1705\,MeV \;(I=0), 1780\,MeV \;(I=1)$ with very large width. 
If the data on narrow low lying pentaquarks are not confirmed, then these earlier predictions have
more chances to be correct.

In the context of chiral soliton model the $\{\overline{10}\}$ and $\{27\}$ -plets of exotic baryons 
were mentioned first in \cite{mc}, without any mass
 estimates, however. 
Rough estimate within 
the "toy" model, $M_{\overline{10}} \,-\,M_8 \simeq 600\,MeV$,
was made a year later in \cite{bieden}. A resonance-like behaviour of $KN$ scattering phase 
shift in $\Theta$ channel was obtained in 
\cite{km} in a version of Skyrme model (in the limit $M_K=M_\pi$)\footnote{Recently this result has
been criticized in \cite{ikor}.}.

Numerical estimate $M_\Theta \simeq 1530\;MeV $ was obtained first by M.Praszalowicz \cite{mic}.
The mass splittings within
 octet and decuplet of baryons have not been described 
satisfactorily with parameters of the model accepted that time, but in the "flexible" approach
proposed in \cite{mic} 8 masses of octet and decuplet of baryons were fitted with Skyrme model
motivated mass formula, depending on 4 parameters, defined from this fit. Central value of the 
mass of antidecuplet
 was found equal to $1706\,MeV$.

Extension of quantization condition \cite{g} to "exotic" case was made in \cite{vk} where
masses of exotic baryonic systems ($B$ arbitrary, $N_c=3$) were estimated as function
of the number of additional quark-antiquark pairs m: 
$\Delta E \sim m/\Theta_K,\;m^2/\Theta_K$. 
It was neither mass splittings estimates inside 
of multiplets, nor calculations of masses of 
particular exotic baryons in \cite{vk}, although
it was shown that baryonic states with additional quark-antiquark pairs appear quite naturally
within chiral soliton approach as $SU(3)$ rotational excitations.
 Recently more general consideration
of such states has been performed in \cite{jenm} for arbitrary numbers of colors and flavors.

First calculation with configuration mixing due to flavor symmetry breaking $(m_K \neq m_\pi)$ was
made by H.Walliser in \cite{hw} where
mass splittings within the octet and decuplet of baryons were well described, and estimate obtained
$M_\Theta \simeq 1660 \;MeV$. "Strange" or kaonic inertia $\Theta_K$ which governs the mass splitting
between exotic and nonexotic baryon multiplets was underestimated in this work, 
as it is clear now (see below).

The estimate $M_\Theta \simeq 1530\;MeV$, coinciding with \cite{mic}, and first estimate of
 the
width, $\Gamma_\Theta < 30\,MeV$ were made later by D.Diakonov, V.Petrov and M.Polyakov \cite{dpp} 
in a variant of quark-soliton model. 
It was "a luck", as stated much later by same authors: 
mass splitting inside of 
$\overline{10}$
 was obtained equal to $540\;MeV$, greater than for decuplet of baryons, and it was 
supposed that 
resonance $N^*(1710)\in \,\{\overline{10}\}$, i.e. it is the nonstrange 
component of antidecuplet. The
above mass value of $\Theta^+$ was a result of subtraction,  $1530 = 1710\,-\;540/3$.
However, the paper \cite{dpp}, 
being in right 
direction, stimulated successful (as we hope still!) 
searches for $\Theta^+$ in RCNP (Japan) \cite{1} and 
ITEP (Russia) \cite{2}. 

Skyrme-type model with vibrational modes included was studied in details first by H.Weigel \cite{wei} 
with a result
$M_\Theta \simeq 1570\;MeV,\;\; \Gamma_\Theta \sim \,70\,MeV$. An inconsistency  in 
width estimate of \cite{dpp} was noted here. 

\begin{center}
\underline{Developements after $\Theta^+$ discovery.}
\end{center}
After discovery of pentaquarks there appeared big amount of papers on this subject which develop
theoretical ideas in different directions: within chiral soliton models 
\cite{wk,mic2,bfk,mic3,wuma,ell,wei1,tram,mix} 
and many
 other; phenomenological correlated quark models \cite{kl,jw,clo,clodu,wil} and other, critical 
discussion by F.Close can be found in \cite{clo2};
QCD sum rules \cite{io,dor}; by means of lattice calculations \cite{csi,ssa}, etc. It is not 
possible 
to describe all of them within restricted framework of present paper 
(reviews of that topic from
 different sides can be found, e.g. in \cite{jm,ufk}).
Quite sound criticism concerning rigid rotator quantization within chiral soliton models was 
developed in \cite{coh,coh3,ikor}, but it
should be kept in mind that the drawbacks of soliton approach should be compared with
uncertainties and drawbacks of other models \footnote{It is stated in conclusions of second of
papers \cite{coh}: "this paper {\it does not} show that rigid-rotor quantization is necessarily invalid but rather that
it is not justified due to large $N_c$ QCD. It remains possible that it is justified due to some
other reason". We agree with this rather optimistic conclusion.}.
There is no regular way of solving relativistic many-body problem to find bound states or
resonances in 3-,5-, etc. quark system, and the chiral soliton approach, in spite of its drawbacks,
provides a way to circumwent some of difficulties. 
The correlated quark models, diquark-triquark model \cite{kl}, or
diquark-diquark-antiquark model \cite{jw}, being interesting and predictive, contain certain,
and very substantial
, phenomenological assumptions.
\section{Topological soliton model}
In spite of some uncertainties and discrepances between different authors,
the chiral soliton approach provided predictions for the masses of exotic states
near the value observed later, considerably more near than quark or quark-bag models 
made up to that time.
Here I will be restricted with this model, mainly. Situation is somewhat paradoxical:
it is easier to estimate masses of exotic states within chiral soliton models,
whereas interpretation is more convenient in terms of simplified quark model.

The topological soliton model is very elegant and attractive (to authors opinion) since it allows to
consider the families ($SU(3)$-multiplets) of baryons, nonexotic and exotic, in unique way.
At the same time, as mentioned in literature, see e.g. \cite{jw}, the apparent drawback
is that this approach does not predict anything about exotics in meson sector.
In these models the baryons and baryonic systems appear as
classical configurations of chiral ("pionic" in simplest $SU(2)$ version) fields
which are characterized by the topological or winding number identified with
the baryon number of the system \cite{sk}. This baryon number is the 4-th component
of the Noether current generated by the Wess-Zumino term in the action written in a
compact form by Witten \cite{wit}, I shall not reproduce it here. In other words, 
the B-number is degree of the map $R^{3} \to SU(2)$, or $R^3 \to S^3$, since $SU(2)$
is homeomorphic to 3-dimensional sphere $S^3$:
\be
\label{wind}
B={-1\over 2\pi^2}\int s_f^2s_\alpha I\biggl[{(f,\alpha,\beta)\over (x,y,z)}\biggr]d^3r 
\ee
where functions $f,\alpha,\beta$, describing $SU(2)$ skyrmion, define the direction of unit vector 
$\vec{n}$ on 3-dimensional sphere $S^3$ and $I[(f,\alpha,\beta)/(x,y,z)]$ is Jacobian of 
corresponding transformation.
More details can be found, e.g. in \cite{wit,wk,ufk}, see also (\ref{L}) below. 
It is important that the number of dimensions
of the ordinary space, equal to $3$, coincides with the number of degrees of freedom (or generators)
of the $SU(2)$ group, and this makes possible the mapping ordinary space onto isospace.
This can be an explanation why the isospin symmetry takes place in hadronic world.

The effective chiral lagrangian describing low-energy phenomena can be obtained from underlying QCD 
by means of special procedure of
bosonization \cite{de,bal,aa}, it contains infinite number of terms - powers of chiral derivatives, 
and, as it is believed, is equivalent to underlying QCD. Many known features of low energy meson-meson
and meson-baryon interactions found explanation within this effective theory. In 
soliton models the truncated lagrangian is used as starting point - few first
terms of this expansion are taken into account to insure solitons stabilization by 4-th, sometimes
6-th order term in chiral derivatives \cite{aj}, and to make evaluations technically possible. 
Further progress in this direction is discussed in \cite{marl,marl2}. 

The Lagrangian density of the model in its minimal form is
\be \label{L}
L = -{F_\pi^2\over 16} Tr (l_\mu l_\mu) +{1\over 32\,e^2} Tr [l_\mu l_\nu]^2 +{F_\pi^2m_\pi^2\over 16}
Tr (U +U^\dagger -2),
\ee
where $m_\pi,\,F_\pi$ are pion mass and decay constant taken from experiment, $e$ is the Skyrme
parameter defining the weight of the 4-th order term, stabilizing the soliton, $e\sim 4$, according
to most of latest estimates, which allows to describe the mass splittings of octet and decuplet of 
baryons; $l_\mu = \partial_\mu U U^\dagger$, $U$ is unitary matrix incorporating the chiral fields,
in $SU(2)$ case $U= c_f +i \vec{n}\vec{\tau}$ and 3 components of the unit vector $\vec{n}$ are 
$s_\alpha c_\beta,\,s_\alpha s_\beta,\, c_\alpha$, see Eq. (\ref{wind}) above. For each value of 
baryon number, one should find the classical field configuration of minimal
energy (mass) - this is done often by means of variational minimization numerical codes. For $B=1$
configuration of minimal energy is of so called "hedgehog" type, where chiral field at each space
point can be directed along radius vector drawn from center of soliton ($\vec{n}=\vec{r}/r$, 
for $B=2$ it has torus-like
 form, for $B=3$ it has topology of tetrahedron, etc.
The relation takes place for configurations of minimal energy between second order term, 
fourth order term and the mass term $(M.t.)$ contributions to classical mass of solitons 
(known as Derrick theorem)
\be \label{drc}
M^{(2)} + 3 M.t. = M^{(4)}.
\ee

The next step is the quantization of these configurations to get spectrum of states with definite 
quantum numbers, isospin $I$, strangeness $S$ or hyperchrge $Y$.
 At this point the flavor 
symmetry breaking $(FSB)$ terms in the Lagrangian are important which define the mass splittings
within $SU(3)$ multiplets of baryons or baryonic systems. Usually they are taken in the form
\be \label{fsb}
L_{FSB} = {F_K^2m_K^2-F_\pi^2m_\pi^2\over 24} Tr (1-\sqrt{3}\lambda_8)(U+U^\dagger -2)-
{F_K^2-F_\pi^2\over 48} Tr (1-\sqrt{3}\lambda_8)(Ul_\mu l_\mu + l_\mu l_\mu U^\dagger )
\ee

In the collective coordinates quantization procedure \cite{anw,g} one introduces the angular
velocities of rotation of skyrmion in the $SU(3)$ configuration space, 
$\omega_k$, $k=1,...8$: $A^\dagger(t)\dot{A}(t) = -i \omega_k\lambda_k/2,\;
\lambda_k$ being Gell-Mann matrices, the collective coordinates matrix $A(t)$
is written usually in the form $A=A_{SU2}\,exp(i\nu\lambda_4)A'_{SU2}\,
exp(i\rho \lambda_8/\sqrt{3})$.
The Wess-Zumino term contribution into lagrangian can be calculated explicitly for this ansatz,
$L_{WZ}=-\omega_8 N_cB/2\sqrt{3}$, and so called "right" hypercharge, or hypercharge in the body-fixed
system equals $Y_R=-2\partial L/\partial \omega_8/\sqrt{3} = N_cB/3$. For any $SU(3)$ multiplet 
$(p,q)$ the maximal hypercharge $Y_{max}=(p+2q)/3$, and obviously, inequality should be fulfilled
$p+2q \geq N_cB$, or
\be
\label{m}
p+2q = 3(B+m) \ee
for $N_c=3$, with $m$ positive integer. States with $m=0$ can be called, naturally, minimal 
multiplets. For $B=1$ they are well known octet $(1,1)$ and decuplet $(3,0)$ \cite{g}.

States with $m=1$ should contain at least one $q\bar{q}$ pair, since they contain the $S=+1,\, Y=2$
hyperon. They are pentaquarks antidecuplet $(p,q)=(0,3)$, $27$-plet $(2,2)$, $35$-plet $(4,1)$.
The pentaquark multiplets are presented in {\bf Figure}.
\begin{figure}[h]
\label{multiplet}
\setlength{\unitlength}{1.cm}
\begin{flushleft}
\begin{picture}(12,14)
\put(3,11){\vector(1,0){2.5}}
\put(3,11){\vector(0,1){3}}
\put(2.6,13.7){$Y$}
\put(5.2,10.6){$I_3$}
\put(2,8.3){$\{\overline {10}\}\, J=1/2$}
\put(3.1,13.1){$\Theta^+$}

\put(1.4,13.0){$(uudd\bar{s})$}
\put(3,13){\circle*{0.2}}
\put(2.5,12){\circle*{0.1}}
\put(3.5,12){\circle*{0.1}}
\put(2,11){\circle*{0.1}}
\put(3,11){\circle*{0.1}}
\put(4,11){\circle*{0.1}}
\put(1.5,10){\circle*{0.2}}
\put(.0,10){$(ddss\bar{u})$}
\put(2.5,10){\circle*{0.18}}
\put(3.5,10){\circle*{0.18}}
\put(4.5,10){\circle*{0.18}}

\put(1.5,10){\line(1,0){3}}
\put(1.5,10){\line(1,2){1.5}}
\put(4.5,10){\line(-1,2){1.5}}

\put(4.6,10){$(uuss\bar{d})$}

\put(10,11){\vector(1,0){3}}
\put(10,11){\vector(0,1){3}}
\put(9.6,13.7){$Y$}
\put(12.7,10.6){$I_3$}
\put(9,8.3){$\{27\}\, J=3/2;\,1/2$}
\put(8.7,13.1){$\Theta$*$^0$}
\put(10.1,13.1){$\Theta$*$^+$}
\put(11.1,13.1){$\Theta$*$^{++}$}

\put(7.3,12.8){$(uddd\bar{s})$}
\put(9,13){\circle*{0.18}}
\put(10,13){\circle*{0.18}}
\put(11,13){\circle*{0.18}}
\put(11.5,12.8){$(uuud\bar{s})$}

\put(8.5,12){\circle*{0.1}}
\put(9.5,12){\circle*{0.1}}
\put(9.5,12){\circle {0.2}}
\put(10.5,12){\circle*{0.1}}
\put(10.5,12){\circle {0.2}}
\put(11.5,12){\circle*{0.1}}

\put(6.4,11){$(ddds\bar{u})$}
\put(8,11){\circle*{0.18}}
\put(9,11){\circle*{0.1}}
\put(10,11){\circle*{0.1}}
\put(11,11){\circle*{0.1}}
\put(12,11){\circle*{0.18}}
\put(9,11){\circle {0.2}}
\put(10,11){\circle {0.2}}
\put(11,11){\circle {0.2}}
\put(10,11){\circle {0.3}}
\put(12.1,11.1){$(uuus\bar{d})$}

\put(6.9,10){$(ddss\bar{u})$}
\put(8.5,10){\circle*{0.18}}
\put(9.5,10){\circle*{0.1}}
\put(9.5,10){\circle {0.2}}
\put(10.5,10){\circle*{0.1}}
\put(10.5,10){\circle {0.2}}
\put(11.5,10){\circle*{0.18}}
\put(11.7,10){$(uuss\bar{d})$}

\put(7.4,9){$(dsss\bar{u})$}
\put(9,9){\circle*{0.18}}
\put(10,9){\circle*{0.18}}
\put(11,9){\circle*{0.18}}
\put(11.2,9){$(usss\bar{d})$}

\put(8,11){\line(1,2){1}}
\put(8,11){\line(1,-2){1}}
\put(9,13){\line(1,0){2}}
\put(9,9){\line(1,0){2}}
\put(12,11){\line(-1,2){1}}
\put(12,11){\line(-1,-2){1}}


\put(7,5){\vector(1,0){4}}
\put(7,5){\vector(0,1){3}}
\put(6.6,7.7){$Y$}
\put(10.7,4.6){$I_3$}
\put(6,0.9){$\{35\}\, J=5/2;3/2$}

\put(3.5,7){$(dddd\bar{s})$}
\put(5,7){\circle*{0.18}}
\put(6,7){\circle*{0.18}}
\put(7,7){\circle*{0.18}}
\put(8,7){\circle*{0.18}}

\put(9,7){\circle*{0.18}}

\put(9.1,7){$(uuuu\bar{s})$}

\put(3.,6){$(dddd\bar{u})$}
\put(4.5,6){\circle*{0.18}}
\put(5.5,6){\circle*{0.1}}
\put(5.5,6){\circle {0.2}}
\put(6.5,6){\circle*{0.1}}
\put(6.5,6){\circle {0.2}}
\put(7.5,6){\circle*{0.1}}
\put(7.5,6){\circle {0.2}}
\put(8.5,6){\circle*{0.1}}
\put(8.5,6){\circle {0.2}}
\put(9.5,6){\circle*{0.18}}

\put(9.6,6){$(uuuu\bar{d})$}

\put(3.5,5){$(ddds\bar{u})$}
\put(5,5){\circle*{0.18}}
\put(6,5){\circle*{0.1}}
\put(7,5){\circle*{0.1}}
\put(8,5){\circle*{0.1}}
\put(9,5){\circle*{0.18}}
\put(6,5){\circle {0.2}}
\put(7,5){\circle {0.2}}
\put(8,5){\circle {0.2}}
\put(9.1,5.1){$(uuus\bar{d})$}

\put(4.0,4){$(ddss\bar{u})$}
\put(5.5,4){\circle*{0.18}}
\put(6.5,4){\circle*{0.1}}

\put(6.5,4){\circle {0.2}}
\put(7.5,4){\circle*{0.1}}
\put(7.5,4){\circle {0.2}}
\put(8.5,4){\circle*{0.18}}
\put(8.6,4){$(uuss\bar{d})$}

\put(4.5,3){$(dsss\bar{u})$}
\put(6,3){\circle*{0.18}}
\put(7,3){\circle*{0.1}}
\put(7,3){\circle {0.2}}
\put(8,3){\circle*{0.18}}
\put(8.1,3){$(usss\bar{d})$}

\put(5.0,2){$(ssss\bar{u})$}
\put(6.5,2){\circle*{0.18}}
\put(7.5,2){\circle*{0.18}}
\put(7.6,2){$(ssss\bar{d})$}

\put(4.5,6){\line(1,2){0.5}}
\put(4.5,6){\line(1,-2){2}}
\put(5,7){\line(1,0){4}}
\put(6.5,2){\line(1,0){1}}
\put(9.5,6){\line(-1,2){0.5}}
\put(9.5,6){\line(-1,-2){2}}


\end{picture}
\vglue 0.1cm
\caption{\tenrm The $I_3-Y$ diagrams for the multiplets of pentaquarks, $B=1,\;m=1$.
Large full circles show the exotic states, smaller - the cryptoexotic states
which can mix with nonexotic states from octet and decuplet. Manifestly exotic components
of pentaquarks satisfy the relation $I=(5+S)/2$ for strangeness $S \leq 0$ and have unique 
quark contents shown in this figure.}

\end{flushleft}
\end{figure}
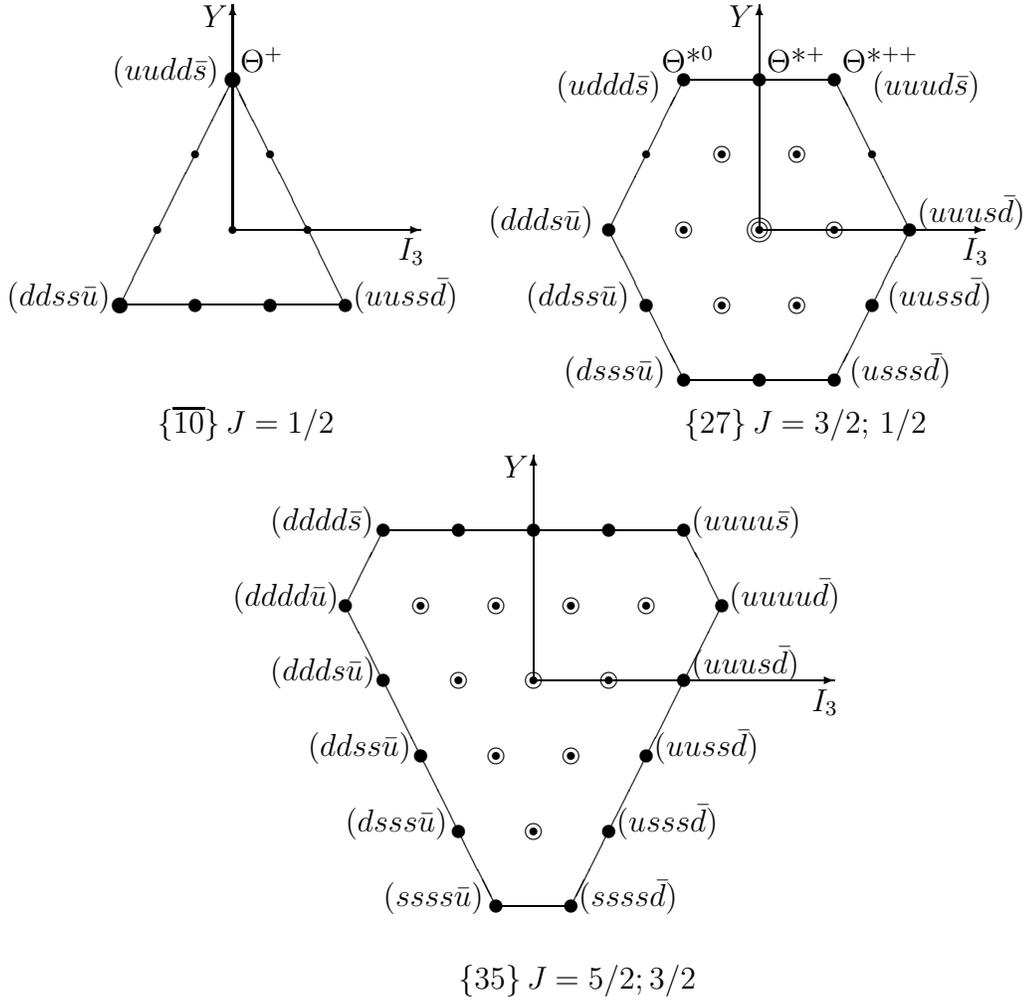
The minimal value of hypercharge is $Y_{min}=-(2p+q)/3$, the maximal isospin
$I_{max}=(p+q)/2$ at $Y=(p-q)/3$. Such multiplets as $\{27\}$, $\{35\}$
for $m=1$ and multiplets for $m=2$
in their internal points contain 2 or more states 
(shown by double or triple circles in {\bf Figure}). The $28$-plet ($6,\,0$)
should contain at least two quark-antiquark pairs, as it follows from analyses of its strangeness
and isospin content \cite{ufk}, so, it is septuquark (or heptaquark)
,
although it has $m=1$, and it is not shown here by this reason.

\section{The mass formula for the rigid rotator}
The lagrangian describing baryons or baryonic system is quadratic form in angular velocities
defined above, with momenta of inertia, isotopical (pionic) $\Theta_\pi$ 
and flavor, or kaonic $\Theta_K$ as coefficients \cite{g}:
\be
\label{Lrot}
L_{rot} = {1\over 2}\Theta_\pi (\omega_1^2+\omega_2^2+\omega_3^2) + 
{1\over 2}\Theta_K (\omega_4^2+...+\omega_7^2) - {N_cB \over 2\sqrt{3}} 
\omega_8. \ee
The expressions for these moments of inertia as functions of skyrmion profile
are well known \cite{anw,g} and presented in many papers, see e.g. \cite{ufk}. The 
quantization condition (\ref{m}) discussed above follows
from the presence of linear in angular velocity $\omega_8$ term in $(3)$
originated from the Wess-Zumino-Witten term in the action of the model 
\cite{wit,g}. 

The hamiltonian of the model can be obtained from $(3)$ by means of canonical 
quantization procedure \cite{g}:
\be
\label{ham}
 H = M_{cl} + {1\over 2\Theta_\pi} \vec{R}^2 + {1\over 2\Theta_K} 
\biggl[C_2(SU_3) -\vec{R}^2 -{N_c^2B^2\over 12} \biggr], \ee
where the second order Casimir operator for the $SU(3)$ group, 
$C_2(SU_3)=\sum_{a=1}^8 R_a^2$, with eigenvalues for the  $(p,q)$ multiplets
$C_2(SU_3)_{p,q}=(p^2+pq+q^2)/3 +p+q, $
 for the $SU(2)$ group,
$C_2(SU2)=\vec{R}^2 =R_1^2+R^2_2+R^2_3= J(J+1) = I_R(I_R+1)$.

The operators $R_\alpha = \partial L/\partial\omega_\alpha$ satisfy definite
commutation relations which are generalization of the angular momentum 
commutation relations to the $SU(3)$ case \cite{g}. Evidently, the linear in
$\omega$ terms in lagrangian (\ref{Lrot}) are cancelled in hamiltonian (\ref{ham}).
The equality of angular momentum (spin) $J$ and the so called right or body 
fixed isospin $I_R$ used in (\ref{ham}) takes place only for configurations of the
"hedgehog" type when usual space and isospace rotations are equivalent. This
equality is absent for configurations which provide the minimum of classical
energy for greater baryon numbers, $B\geq 2$ \cite{vkax}.

\begin{center}
\begin{tabular}{|l|l|l|l|l|l|l|}
\hline

$(p,q)$& $N(p,q)$         &m &$C_2(SU_3)$&$J=I_R$ &$K(J_{max})$&$K(J_{max}-1)$\\
\hline
$(1,1)$&$\{8\}$            &0 & 3        &1/2           &3/2 &\\
$(3,0)$&$\{10\}$           &0 & 6        &3/2           &3/2 &\\

\hline
$(0,3)$&$\{\overline{10}\}$ &1&6         &1/2           &3/2+3&\\
$(2,2)$&$\{27\}$            &1&8         &3/2; 1/2      &3/2+2& 3/2+5\\
$(4,1)$&$\{35\}$            &1&12        &5/2; 3/2      &3/2+1& 3/2+6\\
$(6,0)$&$\{28\}$            &1&18        &5/2           &3/2+7&\\
\hline
$(1,4)$&$\{\overline{35}\}$ &2&12        &3/2; 1/2      &3/2+6& 3/2+9\\
$(3,3)$&$\{64\}$            &2&15        &5/2; 3/2; 1/2 &3/2+4& 3/2+9\\
$(5,2)$&$\{81\}$            &2&20        &7/2; 5/2; 3/2 &3/2+2& 3/2+9 \\
$(7,1)$&$\{80\}$            &2&27        &7/2; 5/2      &3/2+9& 3/2+16\\
$(9,0)$&$\{55\}$            &2&36        &7/2           &3/2+18& \\
\hline
\end{tabular}
\end{center}

{\bf Table 1.}{\tenrm The values of $N(p,q)$, Casimir operator $C_2(SU_3)$, spin
$J=I_R$, coefficient $K$ for first two values of $J$ for minimal $(m=0)$ and 
nonminimal $(m=1,\;2)$ multiplets of baryons.}
\vglue 0.1cm
For minimal multiplets $(m=0)$ the right isospin $I_R=p/2$, and it is easy to
check that coefficient of $1/2\Theta_K$ in (\ref{ham}) equals to
\be
\label{K}
K=\,C_2(SU_3)-\vec{R}^2-N_C^2B^2/12 \,=\,N_CB/2, \ee
for arbitrary $N_C$ \footnote{For the number of colors $N_C$ 
different from 3 the minimal multiplets for baryons differ from octet 
and decuplet. They have $[p,q]=[1,(N_c-1)/2],\; [3,(N_c-3)/2],...,\,[N_c,0]$. There are totally
$(N_c+1)/2$ multiplets of baryons with the interval $\Delta Y$ for each of them increasing from
$(N_c+1)/2$ to $N_c$, see Appendix. It is an artifact
of large $N_c$ approximation that besides the states existing in real world with $N_c=3$, 
the spurious states appear, and the number of these states is infinite as $N_c\to \infty$.}.
So, $K$ is the same for all multiplets with $m=0$ \cite{vk}, see {\bf Table 1} -
the property known long ago for the $B=1$ case \cite{g}.
For nonminimal multiplets there are additional contributions to the energy
proportional to $m/\Theta_K$ and $m^2/\Theta_K$, according to (\ref{ham}).
The following expression was obtained for the energy surplus due to addition of $m$ quark-antiquark
pairs (formula (6) of \cite{vk}):
\be
\label{old}
\delta E_{rot} = m \bigl[(3B/2 +1 +m-N)/(2\Theta_K) + (2N+1-m)/(2\Theta_\pi)\bigr]
\ee
with "right isospin" $N=(p+m)/2$.
It means 
that in the framework of chiral soliton approach the "weight" of quark-
antiquark pair is defined by parameter $1/\Theta_K$, and this property of 
such models deserves better understanding.

{\bf Table 1} was presented previously in \cite{ufk}, here we need it also to estimate 
the mass difference of partners of lowest baryon states with same quantum numbers except
spin. It follows from Table 1 that for each nonzero $m$ the coefficient 
$K(J_{max})$ decreases with increasing $N(p,q)$, e.g. $K_{5/2}(35)\,<K_{3/2}(27)
\,<\,K_{1/2}(\overline{10})$. The following differences of the rotation energy
can be obtained easily:
\be
\label{d108}
M_{10} - M_{8} = {3\over 2\Theta_\pi}, \qquad M_{\overline{10}} -M_{8} = {3\over 2\Theta_K}, \ee
obtained in \cite{g,dpp},
\be
\label{d2710}
M_{27,J=3/2}- M_{10} = {1\over \Theta_K}, \quad M_{27,J=3/2} - M_{\overline{10}} = 
{3\over 2\Theta_\pi} - {1\over 2\Theta_K},
\ee
\be\label{278}
M_{27,J=3/2}- M_{8} = {1\over \Theta_K} +{3\over 2\Theta_\pi} ,
\ee
which follows also from (\ref{old}) at $B=1,\;m=1$,
\be
\label{d3527}
M_{35,J=5/2} - M_{27,J=3/2} = {5\over 2\Theta_\pi} - {1\over 2\Theta_K}.
\ee
According to these relations, the mass difference between decuplet and $27$-plet is 
$1.5$ times smaller than between octet and antidecuplet, by this reason the mixing
between corresponding components of decuplet and $27$-plet is large and not negligible,
although it was neglected in \cite{dpp,dp}, e.g.
If the relation took place $\Theta_K \ll \Theta_\pi$, then $\{27\}$-plet would 
be lighter than antidecuplet, and $\{35\}$-plet would be lighter than $\{27\}$-plet.
In realistic case $\Theta_K$ is approximately twice smaller than $\Theta_\pi$
(see {\bf Table 2}, next section), and therefore the components of antidecuplet are lighter
than components of $\{27\}$ with same values of strangeness.
Beginning with some values of $N(p,q)$ coefficient $K$ increases strongly, as can
be seen from {\bf Table 1}, and this corresponds to the increase of the number 
of quark-antiquark pairs by another unity. The states with $J\,<\,J_{max}$ have
the energy considerably greater than that of $J_{max}$ states, by this reason 
they could contain also greater amount of $q\bar{q}$-pairs.

The states with $m=2$ (second floor of this building) have considerably higher energy than
states with $m=1$, and this difference defines the scale of mass splitting between partners
with higher values of spin. It can be noted from {\bf Table 1} that the $35$-plet with 
$J=3/2$ has exactly the same rotation energy as $\overline{35}$-plet, $(m=2)$ with $J=3/2$:
$K=3/2+6$ in both cases (this degeneracy was noted first by H.Walliser). It is a clear hint
that states with same values of $(p,q)$ but different spin should have different
quark contents.

The formula (\ref{ham}) is obtained in the rigid rotator approximation which is 
valid if the profile function of the skyrmion and therefore its dimensions and 
other properties are not changed when it is rotated in the configuration space 
(see next section and, e.g.
 discussion in \cite{ufk}).
\section{Spectrum of baryonic states}
Expressions (\ref{ham}-\ref{d3527}) and numbers given in {\bf Table 1} are sufficient 
to calculate the spectrum of baryons without mass splitting inside of $SU(3)$-
multiplets, as it was made e.g. in \cite{bieden,vk}.
\subsection{The rigid rotator model}

The mass splitting due to the presence of flavor symmetry breaking terms in the lagrangian (\ref{fsb})
plays a 
very substantial role in the spectrum of baryon states.
The corresponding contribution into hamiltonian can be written in simple form \cite{schw,hw,wk}:
\be \label{HSB}
H_{SB}=\frac{1-D_{88}^{(8)}}{2}\Gamma_{SB}={3\over 4} s^2_\nu\,\Gamma_{SB} \ee
where the $SU(3)$ rotation function $D_{88}^8(\nu) =1-3s^2_\nu/2$,
\be \label{GSB}
\Gamma_{SB}={2\over 3}\Biggl[\Biggl({F_K^2\over F_\pi^2}m_K^2 -m_\pi^2\Biggr)
\Sigma +(F_K^2-F_\pi^2)\tilde{\Sigma}\Biggr] \ee

\be \label{Sig}
 \Sigma = \frac{F_{\pi}^2}{2} \int (1-c_f) d^3\vec{r}, \qquad 
\tilde{\Sigma}= {1\over 4}\int c_f \Biggl(f'^2+{2s_f^2\over r^2}\biggr)d^3r,
\ee
kaon and pion masses $m_K,\;m_\pi$ as well as $F_K,\,F_\pi$ are taken from experiment.
At large number of colors $\Gamma_{SB} \sim N_c$, both the soliton mass and the total mass splitting
of $SU(3)$ multiplets also are $\sim N_c$. Individual mass splittings within multiplets are
of the order of $N_c^0\sim 1$, since the change of hypercharge within multiplets $\Delta Y \sim N_c$, 
see footnote in previous section.
The quantity $SC=<s_\nu^2>/2=<1-D_{88}^{(8)}>/3$ averaged over the baryon
$SU(3)$ wave function defines its strangeness content, which allows to establish a bridge between
chiral soliton approach and quark models \footnote{The $\nu$-dependent wave functions in $SU(3)$ 
configuration space are simple in some cases: e.g. for $\Theta^+$-hyperon $\Psi_\Theta (\nu)=
\sqrt{15} sin\,\nu \,cos^2\nu$, for the hyperon $\Omega\in \{10\}$, $\Psi_\Omega =\sqrt{5/2}sin^3\nu$.
They are normalized according to $\int |\Psi_B(\nu)|^2 4\,sin^3\nu cos\nu d\nu =1$. In most cases
the $\nu$-dependent contributions into $|\Psi_B|^2$ consist of several terms.}.
Without configuration 
mixing, i.e. when flavor symmetry breaking terms in the lagrangian are 
considered as small perturbation, $<s_\nu^2>_0$ can be expressed simply in 
terms of the $SU(3)$ Clebsh-Gordan coefficients. The values of $<s_\nu^2>_0$
for the octet, decuplet, antidecuplet and some components of higher multiplets
are presented in {\bf Tables 2, 5}. In this approximation the components of $\{10\}$ 
and $\{\overline{10}\}$ are placed equidistantly, and splittings of decuplet
and antidecuplet are equal.

The spectrum of states with configuration mixing and diagonalization of the
hamiltonian in the next orders of perturbation theory in $H_{SB}$ is given in 
{\bf Table 2} 
(the code for calculation was kindly presented by H.Walliser).
The calculation results in the Skyrme model with only one adjustable parameter -
Skyrme constant $e$ ($F_\pi=186\,Mev$ - experimentally measured value) are shown
as variants A and B. The values of mass of $\Theta^+$ obtained in this way are close
to the observed mass. The values of $<s_\nu^2>$ become lower when configuration mixing
takes place, and equidistant spacing of components inside of antidecuplet is violated.
For the decuplet of baryons the total mass splitting increases when mixing is included,
but results obtained mimic approximately the equidistant position of its components, see also
discussion in the next section.

It should be stressed here that the chiral soliton approach in its present state
can describe the differences of baryon or multibaryon masses \cite{schw,hw,wk}.
The absolute values of mass are controlled by loop corrections of the order of
$N_C^0\sim 1$ which are estimated now for the case of $B=1$ only \cite{mou}. 
Therefore, the value of nucleon mass in {\bf Table 2} is taken to be equal to 
the observed value.

As it can be seen from Table 2, the agreement with data for pure Skyrme model with 
one parameter is not perfect, but the observed mass of $\Theta^+$ is reproduced
with some reservation. To get more reliable predictions for masses of other
exotic states the more phenomenological approach was used in \cite{wk} where
the observed value $M_\Theta =1.54\,GeV$ was included into the fit, and 
$\Theta_K,\;\Gamma_{SB}$ were the variated parameters (variant C in  Table 2). 
The position of some components of $\{27\}$, $\{35\}$ and $\{\overline{35}\}$-plets is shown 
in Table 2 as well (they are components with largest isospin).

\begin{center}
\begin{tabular}{|l|l|l|l|l|l|}
\hline
   &     & A & B& C &      \\

\hline
$\Theta_\pi \,(GeV^{-1})$ & --- &$6.175 $& $5.556$ & $5.61$ & -    \\
$\Theta_K  \,(GeV^{-1})$   & --- &$2.924 $  & $2.641$ & $2.84$ & -    \\
$\Gamma_{SB} \,\;(GeV) $       & --- & $1.391$  & $1.274 $ & $1.45$ & -    \\
\hline
\hline
$Baryon|N,Y,I,J>$ &$<s_\nu^2>_0$& A & B  & C   & $Data$\\
\hline
$ N\,|8,1,{1\over 2},{1\over 2}>\,input$&$0.467$ &$\,939$ &$\,939$ &$\,939$&$\,939$ \\
$\Lambda\,|8,0,0,1/2>$  &$0.600$ & 1097 & 1082 & 1103   &1116 \\
$\Sigma\,|8,0,1,1/2> $  &$0.733$ & 1205 & 1187 & 1216   &1193 \\
$\Xi  \,|8,-1,1/2,1/2>$ &$0.800$ & 1316 & 1282 & 1332   &1318 \\
\hline
$\Delta \,|10,1,3/2,3/2>$ &$0.583$& 1228 & 1258 & 1253 &1232\\
$\Sigma^*|10,0,1,3/2>$  &$0.667$  & 1359 & 1376 & 1391 &1385\\
$\Xi^* |10,-1,1/2,3/2>$  &$0.750$ & 1488 & 1489 & 1525 &1530\\
$\Omega \,|10,-2,0,3/2> $ &$0.833$& 1611 & 1596 & 1654 &1672\\
\hline
$\Theta^+\,|\overline{10},2,0,1/2>$  &$0.500$& 1521 & 1566 & 1539 &1540 \\
$ N^*\,|\overline{10},1,1/2,1/2>  $  &$0.583$ & 1637 & 1669 & 1661 &1675?\\
$\Sigma^*\,|\overline{10},0,1,1/2>$  &$0.667$ & 1736 & 1756 & 1764 &1770?\\
$\Xi_{3/2}|\overline{10},-1,3/2,1/2>$&$0.750$ & 1758 & 1787 &1786 &1862? \\
\hline
\hline
$\Theta^*_1\,|27,2,1,3/2>$      &$0.571$  & 1648 & 1700 & 1688 &  \\ 
$\Delta^*_1\,|27,1,3/2,3/2>$    &$0.589$  & 1780 & 1809 & 1826 &  \\ 
$\Sigma^*_2\,|27,0,2,3/2>$      &$0.607$  & 1677 & 1728 & 1718 &  \\
$\Xi^{*}_{3/2}|27,-1,3/2,3/2>$  &$0.714$  & 1803 & 1842 & 1850 & 1862? \\
$\Omega^*_1\,|27,-2,1,3/2>$     &$0.821$  & 1935 & 1959 & 1987 &   \\
\hline  
\hline
$\Theta^*_2\,|35,2,2,5/2>$ &$0.708$ & 1982 & 2060 & 2061 &\\ 
$\Delta_{5/2}|35,1,5/2,5/2>$&$0.438$& 1723 & 1816 & 1792 &\\

$\Sigma^*_2\,|35,0,2,5/2>$     &$0.542$& 1844 & 1926 & 1918 &\\
$\Xi^{*}_{3/2}|35,-1,3/2,5/2>$ &$0.646$& 1967 & 2037 & 2046 &\\
$\Omega^{*}_1\,|35,-2, 1,5/2>$ &$0.750$& 2091 & 2149 & 2175 &\\
$ \Gamma \;\, |35,-3,1/2,5/2>$ &$0.854$& 2216 & 2261 & 2306 &\\
\hline
\hline
$\;\; |\overline{35},3,1/2,3/2> $ &$0.562$& 2350 & 2470 & 2412 &\\ 
$\;\; |\overline{35},2,1,3/2>   $ &$0.583$& 2403 & 2513 & 2466 &\\

$\;\; |\overline{35},1,3/2,3/2> $ &$0.604$& 2435 & 2541 & 2501 &\\
$\;\; |\overline{35},0,2,3/2>   $ &$0.625$& 2437 & 2546 & 2502 &\\
$\;\; |\overline{35},-1,5/2,3/2>$ &$0.646$& 2417 & 2534 & 2480 &\\
$\;\; |\overline{35},-2,2,3/2>  $ &$0.792$& 2573 & 2677 & 2643 &\\
\hline
\end{tabular}
\end{center}

{\bf Table 2.} {\tenrm Values of masses of the octet, decuplet, antidecuplet,
manifestly exotic components of higher multiplets, and highest isospin components
of the $\{\overline{35}\}$-plet ($m=2$). 
A: $e=3.96$; B: $e=4.12$; 
C: fit with parameters $\Theta_K,\;\Theta_\pi$ and 
$\Gamma_{SB}$ \cite{wk}, which are shown in the upper 3 lines.}
\\

The experimental value of the mass of the $N^*$ candidate for antidecuplet, $M_{N^*}=1675\,MeV$,
is taken from recent work \cite{kuz}, the value $M_{N^*}\simeq 1680\,MeV$ was obtained somewhat
earlier \cite{araz} in the modified partial wave anlysis of existing data.
The accuracy of numerical calculations of masses given in Table 2 is not better than $\sim 1\%$, and
several figures are presented for convenience of different variants comparison. Small correction of 
numbers for variants A and B is made, in comparison with \cite{ufk}. 
The accuracy of method itself is difficult to
estimate, comparison with other quantization schemes can be useful for this.
The unexpected at first sight fact that the state $\Theta^+\in \{\overline{10}\}$ containing 
strange
 antiquark is lighter than nonstrange component of antidecuplet, $N^*(I=1/2)$
can be easily to understood if we recall that all antidecuplet components contain
$q\bar{q}$ pair: $\Theta^+$ contains 4 light quarks and $\bar{s}$, $N^*$ 
contains 3 light quarks and $s\bar{s}$ pair with some weight, $\Sigma^* \in 
\{\overline{10}\}$ contains $u,d,s$ quarks and $s\bar{s}$, etc.
, see Section 7.

The mass splitting inside of decuplet is influenced essentially by its mixing
with $\{27\}$-plet components \cite{wk}, see {\bf Figure}, which increases this 
splitting considerably - the effect ignored in \cite{dpp}. As a result of this mixing, the lowest
in energy state of $\{10\}$, $\Delta$-isobar, moves considerably towards lower mass, the whole
mass splitting within decuplet increases from $\sim 270\,MeV$ to $350\,-\, 400\,MeV$, but equidistant
position of states remains, roughly, since $\Sigma^*$ and $\Xi^*$ are mixed as well. The 
approximate equidistant position of the components of decuplet is not an argument against
important role of mixing with other multiplets, as it is stated sometimes in literature.

The mixing of 
antidecuplet with the octet of baryons has considerable effect on the position of
$N^*$ and $\Sigma^*(\overline{10})$ - their masses increase, the position of $\Xi^*_{3/2}$ 
is influenced
 by mixing with 
$\{27\}$-plet, $(J=1/2)$, and $\{\overline{35}\}$-plet, and its mass decreases. As a result of mixing, the 
total mass splitting 
of antidecuplet decreases slightly, opposite to the case of decuplet, and
equidistant position of its components is perturbed.
Position of $\Theta^* \in \{27\}$ is influenced by mixing with higher multiplets \cite{wk}, 
the components of $\{35\}$-plet mix mainly with corresponding components of septuquark
$\{64\}$-plet.

The mass of $\Theta$ hyperon is obtained in the interval $1520 - 1560\, MeV$ for the Skyrme 
parameter $e$ between $3.96$ and $4.12$, in the rigid rotator approximation, so the statement made 
in \cite{clo2}
that "all models appear to normalise to some feature and do not naturally explain the low mass
of the orbitally excited pentaquark" does not apply to the simple chiral soliton model which is
$SU(3)$ generalization of the original Skyrme model quantized as rigid rotator.

The flavor symmetry 
breaking in the kaon decay constant, i.e. the fact that $F_K/F_\pi =1.22$ leads to certain increase
of the kaonic moment of inertia and to decrease of the $\Theta$ mass \cite{wk}
\be 
\label{thf}
\Theta_K ={1\over 8} \int (1-c_f) \biggl[ F_K^2 - (F_K^2-F_\pi^2){2-c_f\over 2} s_\nu^2 + 
{1\over e^2} \biggl(f'^2+{2s_f^2\over r^2}\biggr)  \biggr] d^3r,
\ee
where $f(r)$ is the profile function of skyrmion, $f(0)=\pi$ at the origin, and $f(\infty)=0$.
This moment of inertia is maximal when the angle of rotation into strange direction $\nu=0$, see
Table 2, and this corresponds to rigid rotor approximation used previously and in \cite{wk}.
$\Theta_K$ decreases when $\nu$ deviates from $0$, and indeed, the masses of exotic baryons obtained
within soft, or slow rotator approximation, are considerably greater \cite{kss}.

It should be noted that predictions of the mass of $\Xi^*_{3/2}$ made in \cite{wk} half a year before
its observation at CERN \cite{alt} were quite close to the reported value $1862 \,MeV$: it was 
$1786\,MeV$
 for the component of antidecuplet, and $1850\,MeV$ for the $\{27\}$ component, 
variant C of
 {\bf Table 2}.
 Predictions for masses of cryptoexotic components of $\{\overline{10}\}$
and $\{27\}$-plet are clear from {\bf Table 2} as well.

It was stated in the paper \cite{jw} that the spectrum of antidecuplet obtained
 "from correlated
quark picture differs in several dramatic ways from the spectrum predicted by the chiral soliton
model", and "the prediction of light charged exotic $\Xi$'s is the most distinctive signature of
our model". The following comment is necessary here. 
Indeed, it was apparent contradiction of estimates made in \cite{jw} and results
obtained in \cite{dpp} within particular variant of chiral quark-soliton model
where the total splitting of antidecuplet was found equal to $540\,MeV$ \footnote{Later, when 
observation of the resonance $\Xi^{--}$ has been reported in \cite{alt}, the authors 
\cite{dpp} revised their result in \cite{dp} and identified the lowest and the highest masses of 
antidecuplet
with observed values, i.e. they put the total mass splitting of $\overline{10}$ equal to the value
$\sim 324\,MeV$ taken from experiment. Cryptoexotic components of antidecuplet, $N^*$ and $\Sigma^*$,
have been placed within this mass gap. The change of the experimental value of the pion-nucleon 
sigma-term was important for this re-analysis, the latest and largest value of sigma-term was 
obtained in  \cite{pavar}.}. 
As shown in \cite{wk} and above, the total splitting of anti-10
in Skyrme-type model equals to $\Delta_M(\overline{10})= 1.5\;\Gamma_{SB} \Delta_{SC}(\overline{10})
\simeq 270\,MeV$ if configuration mixing is not included, and less than
$\sim 270\,MeV$, about $250\,MeV$ if mixing is taken into account. $\Delta_{SC}(\overline{10})=1/8$ 
is the splitting of strangeness contents within antidecuplet, which can be expressed in terms of 
Clebsch-Gordan coefficients of the $SU(3)$ group. Therefore, there is no such 
dramatic difference between  simplified quark model picture of \cite{jw} and chiral soliton model 
predictions, as it was claimed in \cite{jw}, but results of \cite{wk}, available since April 2004,
have not been considered in \cite{jw}. Another comment is that picture of ideal mixing between
octet and antidecuplet of pentaquark states, proposed in \cite{jw}, can be disturbed by mixing
with ground state octet, as it takes place in chiral soliton approach \cite{wk,mix}. 
Normally, the ground state octet contains admixture of antidecuplet (about $(5-6)\%$ in probability 
for the nucleon,  variant C of {\bf Table 2}),
$\{27\}$-plet ($\sim 3\%$), and smaller amounts of higher multiplets, since the number of 
quark-antiquark pairs is not conserved by strong interactions. 
Further developments of the correlated quark models with diquarks or triquarks are of interest,
also if the announced pentaquark states are not confirmed.

In {\bf Table 3} predictions of the mass of exotic $\Phi/\Xi_{3/2}$ state with strangeness $S=-2$ 
which were made {\it before} experimental evidence for this state \cite{alt}, are presented.
The paper \cite{bfk} repeated the approach of \cite{dpp}, the mass of $\Xi_{3/2} \in \{27\}$ 
shown in  {\bf Table 3} was a new result in comparison with \cite{dpp}. The state observed in \cite{alt}
could belong just to $\{27\}$-plet, another doubly strange state from anti-10 should exist in this 
case, with smaller mass and more narrow than the observed one.
Of course, this discussion becomes irrelevant if the $\Phi/\Xi_{3/2}$ state is not confirmed.
\begin{center}
\begin{tabular}{|l|l|l|l|l|l|l|}
\hline
 &DPP\cite{dpp}&WK\cite{wk}& JW\cite{jw}& BFK\cite{bfk} & P\cite{mic2} &Datum\cite{alt}\\
\hline
$\Phi/\Xi_{3/2}\in \{\overline{10}\}$&2070 &1780-1790 & 1750 & ---  & 1800 & 1862 \\
$\;\;\Xi_{3/2}\in \{27\}       $&--- & 1850      & ---  & 2048  & --- & 1862?\\
\hline
\end{tabular}
\end{center}
{\bf Table 3.} {\tenrm Predictions of the mass of doubly strange hyperon $\Phi/\Xi_{3/2}$ in chronological
order.  The value by M.Praszalowicz, $1800\,MeV$, is taken from figure in \cite{mic2} and is 
approximate by this reason.} \\

The component of $\{35\}$-plet with zero strangeness and $I=J=5/2$ is of special 
interest because it has the smallest strangeness content (or $s_\nu^2$) -
smaller than nucleon and $\Delta$, see {\bf Table 2}. It is the lightest component of $\{35\}$-plet, 
and this remarkable
 property has explanation in simplistic pentaquark model, see Section 7 below. 
As a consequence of isospin conservation by
strong interactions it can decay into $\Delta\pi$, but not to $N\pi$ or $N\rho$.

Generally, the baryon resonances which belong to $\{35\}$-plet cannot be obtained in meson-baryon
interactions, or in some decay into meson (from the octet) and baryon, also from octet, due to
$SU(3)$ invariance of strong interactions, see also \cite{azim}. The components of $\{35\}$ with 
highest isospin which are manifestly exotic, cannot be formed in this way also due to isospin 
invariance of strong interactions, and this is essentially more rigorous prohibition.

The masses of lowest $m=2$ multiplet, $\{\overline{35}\}$-plet, are shown in Table 2 as an example:
this state is not a pentaquark, but septuquark (or heptaquark). The partners of antidecuplet
with spin $J=3/2$ are contained within this multiplet, see section 8.

\subsection{Model of the slow (soft) rotator}
An alternative method of calculation is the soft (or slow) rotator approximation developed
for the case of $B=1$ by Schwesinger and Weigel \cite{schw}, and used
in \cite{kss} to describe the strange dibaryons spectrum. It is supposed within this 
approximation that for each value of the angle $\nu$ it is sufficient time to rearrange the profile 
function under influence of forces due to flavor symmetry breaking terms in the Lagrangian.
Estimates show that for $B=1$ the rigid rotator approximation is better, whereas for $B\geq 2$
the soft rotator becomes more preferable. Indeed, the rotation time in cofiguration space can be
estimated as $\tau_{rot} \sim \pi/\omega$, and the angular velocity 
$\omega \sim \sqrt{C_2(SU3)}/\Theta_K$ for definite $SU(3)$ multiplet, see Table 1. It is difficult 
to estimate the time necessary for rearranging the profile function under
influence of FSB forces, one can state only that it is greater than
time necessary for signal propagation accross skyrmion, $\tau_{sign}\sim 2R_S$. So, we come to
the inequality $\pi \Theta_K \ll 2R_S\sqrt{C_2(SU3)}$ which should be fulfilled for the rigid
rotator approximation being valid. For $B=1$ both sides of this inequality are of the same order of
magnitude, although left side is somewhat smaller.

\begin{center}
\begin{tabular}{|l|l|l|l|}
\hline
\hline
$Baryon|N,Y,I,J>$ &$<s_\nu^2>$& SR & $Data$\\
\hline
$ N\,|8,1,{1\over 2},{1\over 2}>\,input$&$0.314$ &$\;\;939$ &$\;\;939$ \\
$\Lambda\,|8,0,0,1/2>$  &$0.500$ & 1110 &1116 \\
$\Sigma\,|8,0,1,1/2> $  &$0.602$ & 1220 &1193 \\
$\Xi  \,|8,-1,1/2,1/2>$ &$0.740$ & 1320 &1318 \\
\hline
$\Delta \,|10,1,3/2,3/2>$ &$0.315$& 1240 &1232\\
$\Sigma^*|10,0,1,3/2>$    &$0.483$& 1415 &1385\\
$\Xi^* |10,-1,1/2,3/2>$   &$0.650$& 1560 &1530\\
$\Omega \,|10,-2,0,3/2> $ &$0.790$& 1670 &1672\\
\hline
$\Theta\;\;|\overline{10},2,0,1/2>$  &$0.380$& 1737  & 1540 \\
$\Xi_{3/2}|\overline{10},-1,3/2,1/2>$&$0.607$ & 2118 & 1862? \\
\hline
\hline
$\Theta^*_1\,|27,2,1,3/2>$    &$0.416$  & 1840 &  \\ 
$\Sigma^*_2\,|27,0,2,3/2>$    &$0.438$  & 1880 &  \\
$\Xi^{*}_{3/2}|27,-1,3/2,3/2>$&$0.594$  & 2090 &  \\
$\Omega^*_1\,|27,-2,1,3/2>$   &$0.755$  & 2270 &  \\
\hline  
\hline
$\Theta^*_2\,|35,2,2,5/2>$     &$0.464$ & 2180 &\\ 
$\Delta_{5/2}\,|35,1,5/2,5/2>$ &$0.242$ & 1750 &\\ 
$\Sigma^*_2\,|35,0,2,5/2>$     &$0.382$ & 1990 &\\ 
$\Xi^*_{3/2}\,|35,-1,3/2,5/2>$ &$0.528$ & 2190 &\\ 
$\Omega^{*}_1\,|35,-2, 1,5/2>$ &$0.675$ & 2370 &\\
$ \Gamma \;\, |35,-3,1/2,5/2>$ &$0.854$ & 2530 &\\
\hline
\hline
$\;\; |\overline{35},3,1/2,3/2> $ &$0.442$& 2900 & \\ 
$\;\; |\overline{35},-1,5/2,3/2>$ &$0.477$& 3030 & \\
$\;\; |\overline{35},-2,2,3/2>  $ &$0.641$& 3385 & \\
\hline
\end{tabular}
\end{center}

{\bf Table 4.} {\tenrm Values of masses of the octet, decuplet, antidecuplet and manifestly
exotic components of higher multiplets
 within soft rotator (SR) approximation \cite{schw}:
$e=3.46,\; F_K/F_\pi =1.26,$ the values of $<s^2_\nu>$ are calculated with configuration mixing
included.}\\

Thus we see, that rigid rotator is really more preferable
for $B=1$, especially for exotic multiplets (since the value of $C_2(SU3)$ is greater, see Table 1) 
\footnote{If we take into account connection between soliton radius and another (isotopic) moment of 
inertia, $\Theta_I \sim M_S R_S^2/2$, this condition will take the form 
$\Theta_K < \sqrt{\Theta_IC_2(SU3)/M_S}$.}.
With increasing baryon number left side of this inequality grows faster than right side, therefore
slow rotator approximation may become better for greater baryon numbers.
In view of conflicting experimental situation on pentaquarks
observation it makes sense to present the results for pentaquarks spectra within this approximation 
as well.

A natural thing is that the masses of exotic states within soft rotator model are greater 
than in the rigid rotor approximation: the strange, or kaonic inertia $\Theta_K$ becomes smaller
for slow rotator, as explained above, (\ref{thf}). However, it looks somewhat unexpected that increase
of masses is so great, up to $\sim 200\, MeV$. Anyway, for $B=1$ the rigid rotator approximation 
is better justified \cite{ufk}. Strangeness contents of baryon states shown in Table 4 are calculated
with configuration mixing included, and they are considerably smaller than SC calculated in first
order of perturbation theory, Table 2. The slow rotator approximation \cite{schw} deserves more 
attention if the negative results on $\Theta$ observation are confirmed.

All baryonic states considered here are obtained by means of quantization of
soliton rotations in $SU(3)$ configuration space, therefore they have positive
parity. A qualitative discussion of the influence of other (nonzero) modes
- vibration, breathing - as well as references to corresponding papers can be 
found in \cite{wei,wk}. Calculation of baryon spectra with monopole excitation is made in
\cite{wei,wei1}. The realistic situation can be more complicated than 
somewhat simplified picture presented here, since each rotation state can have 
vibrational excitations with characteristic energy of hundreds of $MeV$.
 The resonance $N^*(1440)$
is just the monopole excitation of ground state nucleon \cite{wei,wei1}.

\section{Comments on the $\Theta$ width and large $N_c$ arguments}
If the matrix element of the decay $\Theta^+\to KN$ is written in a form
\be
 M_{\Theta\to KN} = g_{\Theta KN}\bar{u}_N \gamma_5 u_\Theta \phi_K^\dagger
\ee
with $u_N$ and $u_\Theta$ - bispinors of final and initial baryons, then the
 decay width equals to
\be \label{Gamma}
\Gamma_{\Theta\to KN}= \frac{g^2_{\Theta KN}}{8\pi}
\frac{\Delta_M^2-m_K^2}{M^2} p_K^{cm} \simeq 
\frac{g^2_{\Theta KN}}{8\pi}
\frac{\bigl(p_K^{cm}\bigr)^3}{Mm_N} 
\ee
where $\Delta_M=M-m_N, \; M$ is the mass of decaying baryon, $p_K^{cm}\simeq 269\, MeV/c$ if 
$M_\Theta = 1540\, MeV$ - the
kaon momentum in the c.m. frame. For the decay constant we obtain then
$g_{\Theta KN} \simeq 4.4$ if we take the value $\Gamma_{\Theta\to KN}=10\,MeV$
as suggested by experimental data \cite{2},\cite{zeus}. This should be compared 
with pion-nucleon coupling constant $g_{\pi NN}\simeq 13.14$ (according to latest analysis
\cite{arbri} $g^2_{\pi NN}/(4\pi) = 13.75 \pm 0.10$). So, suppression of the decay 
$\Theta\to KN$ takes place, but not 
very large if the width is really close to $10\,MeV$.

Prediction of the widths of baryon resonances is not an intrinsic property of the chiral 
soliton approach - in distinction from spectra of states. Additional assumptions concerning
the form of transition amplitudes are necessary \cite{dpp,mic2}.
Numerical cancellation
in the matrix element of $\Theta$ decay was obtained in \cite{dpp}, and later proved also in 
large $N_c$ limit in chiral quark soliton model, for vanishing dimension of the soliton \cite{mic2}.
It would be difficult, however, to
 explain the width $\Gamma_\Theta \sim 1\,MeV$ or smaller, as 
suggested by scattering data
 \cite{asw,gib}.

The arguments have been presented in the literature \cite{coh}, see also \cite{mic2},
that in large $N_c$ limit and in the case of chiral symmetry, i.e. when $m_\pi=m_K=0$, one should
expect that $\Theta$ width is parametrically greater than width of $\Delta (1232)$ isobar, 
in contradistinction from what is seen experimentally.
As it has been observed long ago, the mass splitting between antidecuplet and octet of baryons
is of the order of $N_c^0\sim 1$, whereas that between decuplet and octet is of the order
of $N_c^{-1}$ due to the difference in rotation energy. To make these conclusions, the 
identification of multiplets in our $N_c=3$ world and hypothetical large $N_c$ world is made in 
definite way, and this identification is not unique in the latter case.

It is known that the artifact of large $N_c$ generalization of the chiral soliton and the quark
models is appearance of multiplets of baryons
which are absent in real $N_c=3$ world \footnote{This is not the only problem. The hypercharge for
arbitrary (but odd) $N_c$ is $Y=N_cB/3 +S$ (\cite{g}, see also \cite{ikor,coh}), and the electric 
charge defined by relation $Q=I_3 +Y/2$ is integer only if $N_c$ is multiple of $3$. Another 
possibility for electric charges was discussed in \cite{abb},  see Appendix.}. 
For nonexotic baryons there are $(N_c+1)/2\,$  $SU(3)$
multiplets, beginning with $[p,q]= [1,(N_c-1)/2]$ which is interpreted as analogue of $N_c=3$ octet.
The next one is the multiplet with $(p,q) = [3, (N_c-3)/2]$ interpreted as "decuplet", and the 
multiplet with largest $p$ is that with $(p,q)=(N_c,0)$. All multiplets 
except first two are usually ignored, even not mentioned. To discuss the large $N_c$ properties
of any particular state, one should first establish correspondence between such state in real world
and in miraculous large $N_c$ world, and the way to do this depends on the principle which is 
taken as leading one.
The $Y=N_c/3$ state within $(1, (N_c-1)/2)$ multiplet not only has the minimal possible for any baryon
isospin (and spin) $I=1/2$, but is also the state maximally antisymmetrized in isospin and spin 
variables. It is natural to consider it as analogue of nucleon by this reason as well.

The state with $Y=1$ from $(3,0)$ multiplet in $N_c=3$ world not only is a state with isospin $I=3/2$,
but it is also a state of maximal symmetry in isospin and spin variables. So, if we take this 
principle of maximal symmetry as a leading one, we should take the state with $Y=N_c/3,\;I=N_c/2$ 
from the multiplet $(p,q)=(N_c,0)$ as analogue of $\Delta$ in the large $N_c$ world. 
The rotation energy of this state
quantized as rigid rotor is very large, $N_c(N_c+2)/(8\Theta_I)\sim N_c$, leading to parametrically 
large width of $"\Delta "$ baryon. The ratio of $"\Delta" - "N"$ mass splitting to $"\Theta"-"N"$
splitting is of the order $\sim N_c$ in this case. Even if not quite convincing, this example shows 
that large $N_c$ argumentation is not without ambiguity because identification of baryons of $N_c=3$ 
and large $N_c$ worlds is a subtle question.

The difference of masses of particular
baryons, e.g. of $\Delta \in \{10\}$ and nucleon from ground state octet contains also some 
contribution due to $FSB$ terms and different values of their strangeness content. Strangeness
content of the nucleon is $SC_N=7/30$, and $SC_\Delta = 7/24$, see {\bf Table 2} and formula (\ref{HSB}).
Strangeness contents of analogues of nucleon in $[1,(N_c-1)/2]$  and "$\Delta$" in
$[3,(N_c-3)/2]$ multiplets are given in Appendix. Their difference is
$SC_\Delta - SC_N = 2(N_c+4)\bigl[1/[(N_c+1)(N_c+9)] - 1/[(N_c+3)(N_c+7)]\bigr]$. At large $N_c$
this difference decreases like $1/N_c^3$, therefore it gives negligible contribution to the
baryons mass differences.

In reality the masses of $\pi$-meson and especially of kaons are not only different from zero,
but even comparable with mass splittings - both masses are formally of the order of $N_c^0\sim 1$. 
The width of $\Theta$ depends on result of cancellation of two quantities, each of them is of the 
order of $N_c^0\sim 1$: $"\Theta \,-\,N"$ mass splitting and kaon mass. For the case of 
$\Theta \to KN$ almost all energy release is absorbed by the kaon mass. 
Therefore, the phase space suppression of the $\Theta$ decay cannot be controlled by $1/N_c$ counting 
arguments only, since finally it depends on subtraction of two quantities of same order 
of magnitude, $N_c^0 \sim 1$, but different nature, at least in our present understanding. The result 
of this subtraction looks occasional,
it could be even negative, thus making the $\Theta$ baryon stable relative to strong interactions.
There are many examples in physics when some quantities of crucial importance
cannot be deduced from general principles \footnote{The particular values of binding
energies of nuclei or nuclear levels responsible for stability or instability of nuclear isotopes, 
and for spectra of photons and neutrinos emitted by stars could be one of such examples.}.

Without any doubt, the width of exotic states is extremely important
and interesting quantity, especially if the width of the order of $1\,MeV$ for $\Theta^+$ 
is confirmed. The cheking of scattering data used in the analyses of \cite{asw,gib} seems to be quite
important, see also \cite{azim}. Critical review of these scattering data was made recently in
\cite{hic2} where necessity of their checking also has been emphasized.

\section{Wave functions of pentaquarks and the masses of strange quark (antiquark)}
Similar to the case of baryons and mesons made of valence quarks (antiquarks), it is convenient
to discuss the properties of new baryon resonances in terms of their quark wave functions (WF).
The question about correspondence of chiral soliton model results and expectations from the quark
models is quite interesting and even thorny.
\subsection{Quark contents of pentaquarks}
The quark contents of wave functions of {\it manifestly} exotic resonances are {\it unique} 
within pentaquark
 approximation, i.e. the number of quarks or antiquarks of definite flavor is 
fixed by their strangeness and isospin \footnote{For pentaquarks manifestly exotic states are 
defined by relation $I=(5+S)/2$ if strangeness $S\leq 0$ (any state with $S>0$ is manifestly 
exotic, as discussed in Introduction).
}.
It is easy to obtain for WF of manifestly exotic components of antidecuplet (see also the {\bf Figure}):
$$ \Psi_\Theta \sim uudd\bar{s}, $$
and for 4 components of exotic $S=-2,\,I=3/2$ state
$$ \Psi_{\Phi/\Xi_{3/2}} \sim ssdd\bar{u};\; ssd(u\bar{u}-d\bar{d})/\sqrt{2};\; 
ssu(d\bar{d}-u\bar{u})/\sqrt{2};\; ssuu\bar{d}. $$
Quark content of cryptoexotic states WF are not unique. Within antidecuplet:
$$\Psi_{N^*} \sim udd\,
[\alpha_- u\bar{u} + \beta_-d\bar{d} +\gamma_- s\bar{s}];\;\;
                  uud\, [\alpha_+ u\bar{u} + \beta_+d\bar{d} +\gamma_+ s\bar{s}], $$
$$\Psi_{\Sigma^*} \sim sdd\,[\mu_- u\bar{u} + \nu_-d\bar{d} +\rho_- s\bar{s}];\;...\;;\;
 sdd\,[\mu_+ u\bar{u} + \nu_+d\bar{d} +\rho_+ s\bar{s}], $$
coefficients $\alpha_-,\;\alpha_+$, etc. depend on the particular variant of the model.\\
E.g., for the model with diquark transforming like flavor anti-triplet, $D_q \sim \bar{3}_F$ 
\cite{jw}
, $\alpha_-=\sqrt{1/3},\; \beta_-=0,\; \gamma_-=\sqrt{2/3}$, etc. Equidistancy within
$\overline{10}$ was obtained in \cite{clo,clodu} for this case.

Within $\{27\}$-plet only the $S=0,\;I=3/2$-state (analogue of $\Delta$-isobar) is cryptoexotic.
The states with $S=+1, I=1$ and state with $S=-1, I=2$ contain one $s$-quark field as depicted in
the {\bf Figure}, and their masses do not differ much by this reason, as it was obtained in chiral
soliton model as well, see Table 2.

Within $\{35\}$-plet {\it all} states of maximal isospin are manifestly exotic and have unique 
quark content.
The state with $S=0,\;I=5/2$ (it can be called $\Delta_{5/2}$) does not contain strange quarks:
$$ \Psi_{\Delta_{5/2}} \sim dddd\bar{u};\;...\;;\; uuuu\bar{d}, $$
neither $s$, nor $\bar{s}$ quarks!
Remarkably, that within chiral soliton model this state has minimal, among all baryons, strangeness 
content ($SC(\Delta_{5/2})\simeq 0.22$), 
and has the lowest (within $\{35\}$-plet) mass, see Table 2 and \cite{wk}.

Evidently, besides flavor antitriplet diquark $D_q \sim \bar{3}_F$ (anti-triplet in color, singlet
$L=0$ state) discussed in this context 
in \cite{jw} and called also "good" diquark \cite{wil}, 
the diquarks $D_q \sim 6_F$ ("bad" diquarks, transforming also like $\bar{3}$ in color, triplet
$L=0$ states) are necessary to form $\{27\}$- and $\{35\}$-plets of 
pentaquarks.

Let us denote $(q_1q_2)$ the flavor symmetric diquark with spin $J=1$ ($\bar{3}_C$ in color,
triplet $L=0$ state).
Then realization of the wave function of $\{27\}$-plet of pentaquarks via diquarks is (we use same 
notation $|N(p,q),Y,I,I_3>$ for the components of multiplets as in {\bf Table 2}, and present the 
states with lowest value of $I_3$):
$$\Psi_{|27,2,1,-1>}\sim (d_1d_2)[d_3u_4]\bar{s}, $$
with $[u_3d_4]=(u_3d_4-d_3u_4)/\sqrt{2}$,
$$ \Psi_{|27,1,3/2,-3/2>} \sim \bigl[-(d_1d_2)[u_3d_4]\bar{u}+ (d_1d_2)[s_3d_4] \bar{s}\bigr]/\sqrt{2}, $$
other components of this $Y=1$ isomultiplet can be obtained easily with the help of isospin raising
$I^+$ operator.
 States with negative strangeness have the wave functions
$$ \Psi_{|27,0,2,-2>} \sim (d_1d_2)[s_3d_4]\bar{u}, $$
$$ \Psi_{|27,-1,3/2,-3/2>} \sim (d_1s_2)[s_3d_4]\bar{u}, $$
$$ \Psi_{|27,-2,1,-1>} \sim (s_1s_2)[s_3d_4]\bar{u}. $$
The components with other projections of isospin $I_3$ are not shown here since they can be obtained
easily.

For the $\{35\}$-plet two flavor-symmetric diquarks $D_{6F}$ are necessary to form the states with
maximal isospin, according to group theoretical equality $6\otimes 6\otimes \bar 3 = \{35\} \oplus
\{27\} \oplus 2\{10\} \oplus \{\overline{10}\} \oplus 2\{8\}$.
 For example, the $S=+1$ $\Theta^*$ state has wave function
$$ \Psi_{|35,2,2,-2>} \sim (d_1d_2)(d_3d_4)\bar{s}, $$
the above mentioned $\Delta_{5/2}$ has
$$ \Psi_{|35,1,5/2,-5/2>} \sim (d_1d_2)(d_3d_4)\bar{u}, $$
etc. States with other isospin projections also can be obtained easily.
The antidecuplet can be made from two symmetric diquarks $D_{6F}$ as well, e.g. its $S=+1$ component 
made
 of two isovector diquarks
 is 
$$\Psi_{\Theta^+}\sim [u_1u_2d_3d_4+ d_1d_2u_3u_4-{1\over 2}(u_1d_2+u_2d_1)(u_3d_4+d_3u_4)]\bar{s},$$
it is expected to have considerably higher energy \cite{wil}.
\subsection{Mass of strange antiquark in different pentaquark multiplets}
Here we shall compare the mass spectrum of baryons obtained within chiral soliton model (CSM) 
with the quark model in pentaquark approximation, which will allow to make some conclusions concerning
masses of strange quarks, antiquarks and diquarks, necessary to fit
the chiral soliton model predictions.
The contribution of strange quark mass $(m_s)$, antiquark mass $(m_{\bar s})$ and strange
diquark mass $(m_{s\bar s})$ to masses of pentaquark states is presented in Table 5, in the lines
below notations of states.

It is easy to see that without configuration mixing (the first lines of numbers in Table 5) 
there is linear dependence of masses on hypercharge of states for antidecuplet;
for $\{27\}$-plet the states with hypercharges $Y=2,\;1,\;0$ belong to one line, and states
with $Y=0,\;-1,\;-2$ - to another line; for $\{35\}$-plet 5 states with hypercharge
from $Y=1$ down to $Y=-3$ are on one line.
Such linear dependence, however, is not specific for CSM only, but is the consequence
of the special way of $SU(3)$-symmetry breaking, when FSB terms in lagrangian are proportional 
to the $D_{88}$ Wigner function, or to hypercharge, which leads to the Gell-Mann - Okubo formula 
\be
\label{gmo}
\Delta M_{FSB} = a [Y^2/4 - I(I+1)] + b Y,
\ee
with $a,\,b$ - some constants different for different $SU(3)$ multiplets.

For antidecuplet the relation between hypercharge and isospin takes place $I=1-Y/2$.
For $\{27\}$-plet the states $|Y,I>=|0,2>,\; |-1,3/2>$ and $|-2,1>$ are on the line $I=Y/2+2$,
the states $|Y,I>=|0,2>,\; |1,3/2>$ and $|2,1>$ belong to the line $I=-Y/2 +2 $.
For the components of $\{35\}$-plet with $Y$ from 1 to -3 similar relation takes place
$I=Y/2+ 2 $. It is easy to see that in all these cases quadratic in $Y$ term in formula (\ref{gmo})
cancels, and linear dependence of the mass on hypercharge
takes place. Similar results have been obtained recently in \cite{oh}.

Less trivial and more informative are some relations for masses of strange quarks/antiquarks 
which follow from comparison with the quark model.
In what follows we shall reserve a possibility that effective masses of strange quark
and antiquark are different, as well as they are different within different $(p,\,q)$ multiplets.
This effect is known already since it takes place for ground states octet and decuplet
of baryons as well: the effective strange quark mass is $189\,MeV$ within octet and $147\,MeV$ within
decuplet, in average \footnote{When hyperfine splitting contributions are included, this difference
between strange quark masses extracted from octet and decuplet, becomes much smaller.}. 

We can easily obtain within pentaquark approximation, ascribing the mass difference of different
components to the strange quark (antiquark) mass, the following relations:
\be
\label{splaten}
\Delta_M(\{\overline{10}\}) = [2m_s - m_{\bar{s}}]_{\{\overline{10}\}}. \ee
Recall that for decuplet 
\be\label{splten}
\Delta_M(\{10\}) = [3m_s]_{\{10\}}, \ee
so, in oversimplified model where $m_s(\{\overline{10}\})=m_{\bar{s}}(\{\overline{10}\})=
m_s(\{10\})$ the first one should be 3 times smaller than $\Delta_M(\{10\}) $.
However, this condradiction becomes much softer and even can dissappear in more refined models
where masses of strange quark and antiquark are different, as well as they are different in
different $SU(3)$ multiplets.

The equality of mass differences between adjacent components of antidecuplet:
$$ {2\over 3}m_{s\bar s}-m_{\bar s}=m_s-{1\over 3}m_{s\bar s} $$
has a consequence that the
mass of $s\bar s$ pairs equals simply to the sum of quark and antiquark masses:
\be \label{ss10} m_{s\bar s} = m_s +m_{\bar s}, \ee
the index $\overline{10}$ is omitted for all masses.
Equality similar to (\ref{ss10}) holds for masses within $\{27\}$-plet as well.
Relations (\ref{splaten},\ref{ss10}) are the only relations which can be obtained for
masses of strange quark and antiquark within antidecuplet, leaving otherwise much freedom
for these masses.
\begin{center}
\begin{tabular}{|l|l|l|l|l|l|}
\hline
$|\overline{10},2,0>$&$|\overline{10},1,1/2>$&$|\overline{10},0,1>$&$|\overline{10},-1,3/2>$& &\\
\hline
$m_{\bar s}$         &$2 m_{s\bar s}/3$   &$m_s+m_{s\bar s}/3$&$2 m_s$& & \\
\hline
564 & 655 & 745 & 836 & &\\
600 & 722 & 825 & 847 & & \\
\hline
\hline
$|27,2,1>$&$|27,1,3/2>$&$|27,0,2>$&$|27,-1,3/2>$&$|27,-2,1>$&\\
\hline
$m_{\bar s}$         &$m_{s\bar s}/2$   &$m_s$&$2 m_s$&$3m_s$  & \\

\hline
733 & 753 & 772 & 889 & 1005 &\\
749 & 887 & 779 & 911 & 1048 &\\
\hline
\hline
$|35,2,2>$&$|35,1,5/2>$&$|35,0,2>$&$|35,-1,3/2>$&$|35,-2,1>$&$|35,-3,1/2>$ \\
\hline
$m_{\bar s}$         &$0 $   &$m_s$&$2 m_s$&$3m_s$ &$4 m_s$ \\
\hline
1152 & 857 & 971 & 1084 & 1197 & 1311 \\
1122 & 853 & 979 & 1107 & 1236 & 1367 \\
\hline
\end{tabular}
\end{center}
{\bf Table 5.} {\tenrm Masses of components of $\{\overline{10}\}$, and components with maximal
isospin for
 $\{27\}, J=3/2$ and $\{35\},J=5/2$ -plets of exotic baryons (in $MeV$, the nucleon mass is 
subtracted). The contribution of strange quarks and antiquarks is written below notation of states,
$m_{s\bar s}$ is the mass of the $s\bar s$ pair. 
The first line of numbers is the result of calculation without configuration 
mixing, the second line - configuration 
mixing included according to \cite{wk}. 
Calculations correspond to variant $C$ of {\bf Table 2}, or case $A$ of paper \cite{wk}: 
$\Theta_K =2.84\,GeV^{-1},\; \Theta_\pi=5.61\,GeV^{-1},\;\Gamma =1.45\,GeV.$ } \\

More information can be obtained for strange quark/antiquark masses within higher exotic multiplets.
Linear dependence of masses of manifestly exotic components of $\{27\}$-plet allows to obtain for
the mass of strange quark $m_s\in \{27\} \simeq 117\,MeV$ (configuration mixing not included).
For $\{27\}$-plet it is also useful to fix the difference of masses between manifestly 
exotic components
$|27,2,1>$ and $|27,0,2>$:
\be \label{spl27}
\Delta_{2-0}(\{27\})= [m_{\bar{s}} - m_s]_{\{27\}} \simeq - 40\,MeV,
\ee
so, strange antiquark within $27$-plet should be lighter than strange quark, according to CSM
results.

Within $35$-plet, it follows from the results for masses of the components with strangeness
$S\leq 0$ that the effective strange quark mass is about $115\,MeV$.
If we ascribe the difference of masses between $Y=2$ and $Y=1$ states of $\{35\}$-plet to the mass of
strange antiquark, we obtain that $m_{\bar{s}}\in \{35\} \simeq 295\,MeV$. 
Strong interactions of the quark $s$ and antiquark $\bar{s}$ are different, 
so no wonder that effective masses of quark and antiquark are different. However,
such big difference between masses of strange antiquarks in $27$ and $35$-plets seems
to be unexpected.

Configuration mixing increases the mass of strange quark within $27$-plet
up to $\sim 135\,MeV$. Within $35$-plet configuration mixing does not change the above numbers
drastically: the effective strange quark mass increases up to $125-130 \,MeV$, and the mass of 
antiquark $\bar{s}$ decreases to $\sim 270\,MeV$.

The effect of configuration mixing is especially important for cryptoexotic components of
antidecuplet ($Y=1\;and\;0$) which mix with similar components of the lowest baryon octet, as a result,
their masses increase. The $\Phi/\Xi_{3/2}$ component is mixed with analogous component $\Xi_{3/2}
\in \{27\}$, and its mass moves to lower value. In summary, after mixing the total mass splitting of
antidecuplet decreases, and equidistant position of states is considerably violated, unlike the case
of decuplet. 
Within $27$-plet, configuration mixing increases the mass of cryptoexotic state $|27,1,3/2>$
considerably (more than by $130\,MeV$). The identification of this state, analogue of 
$\Delta (1232)$ -isobar, is not straihgtforward, see also recent analysis in \cite{azim}.

The 
comparison of masses
 of $\Theta^+ \in \overline{10}$ and $\Theta^* \in 27$ allows to 
conclude that $6_F$ diquark is heavier
than $\bar{3}_F$ diquark by $\sim 120-150\,MeV$, if we ascribe the mass difference of 
$\Theta$-resonances to the mass difference of diquarks. 
It is possible to estimate the mass difference of diquarks more straightforward, in the limit 
$m_K \to 0$. Then we obtain $\Delta M_{(6F - \bar{3}F)} \simeq 3/(2\Theta_\pi)-1/(2\Theta_K) \sim
100\,MeV$.  This is smaller than the estimate given by F.Wilczek in \cite{wil},
$\Delta M_{(6F - \bar{3}F)} \sim 240 - 360\,MeV$.

The comparison of masses of $\Theta^* \in 27$ and $\Theta^* \in 35$, taking into account
the mass difference of strange antiquarks, allows to get for the mass difference of bad and
good diquarks, $\Delta M_{(6F - \bar{3}F)} \sim 200\,MeV$, in better agreement with estimate
of \cite{wil}.

As it follows from the consideration of negative
strangeness components of $\{27\}$- and $\{35\}$-plets, the masses of strange quarks do not differ
considerably within these multiplets, they are close to $130\,MeV$ and do not differ much 
from masses of strange valence quarks within octet and decuplet of baryons.
Quite different, even paradoxical situation takes place for strange antiquark. 
If we take the mass of $s$-quark in
antidecuplet about $(140-150)\,MeV$, as in decuplet, then the effective mass of strange anti-quark 
should be small, not greater than few tens of $MeV$. 
For $\{27\}$-plet we obtain from (\ref{spl27}) that
strange anti-quark is lighter than strange quark by $30-40\,MeV$, and in $\{35\}$-plet the mass of
strange anti-quark is about $270-290\,MeV$, or about $\sim 2$ times greater than mass of strange 
quark within $\{27\}$-plet.
Detalization of the quark models could show is it really possible, or not.

To conclude this section, we note that effective masses of strange quark and especially strange 
antiquark should be different for different $SU(3)$ multiplets, to make possible the link between 
rigid rotator version of chiral soliton and simple quark model. This issue will be considered in
more details elsewhere.

\section{Partners of lowest exotic states with different spin}
The partners of lowest exotic states, i.e. the states with same flavor quantum numbers, isospin, 
strangeness, etc., but different spin have been 
discussed in the literature during latest years, after evidence has been obtained for exotic states
like $\Theta^+$ and $\Phi/\Xi_{3/2}$ \cite{clo,coh3}. Within $CSM$ the equality between spin of baryon states
and so called "right" isospin 
($I_R=1/2$ for antidecuplet) follows from the fact that the lowest $B=1$ classical configuration 
is of 
hedgehog type, and as a result the isospin and space (or spin) rotations are equivalent.
It is not so for the states with greater values of $B$ which have generally different spin and isospin
\cite{vkax}
.
At the same time, within the quark or correlated quark models one could expect existence of 
partners of states, since the spins of quarks and angular momentum of orbital motion can be summed 
providing states
 with different values of spin \cite{clo}. For example, according to \cite{clo} one 
should expect existence of partners of $\Theta^+$ with
 $J^P=3/2^+$ and the mass greater than that of 
lowest states by several tens of $MeV$ \cite{clo}. 
This possibility was considered as an argument against chiral soliton models since it was claimed that 
such states cannot be obtained within $CSM$.

However, careful consideration of multiplets of exotic baryons in the framework of chiral soliton
approach allows to conclude that partners of lowest baryons exist within higher $SU(3)$ multiplets 
of baryons. Some examples are considered here.

The partners of baryon antidecuplet ($J^P=1/2^+$) with $J^P=3/2^+$ exist within $\overline{35}$-plet, 
$(p,\,q)=(1,\,4),$ exoticness $m=2$. The rotational energy of these states is greater than that
of antidecuplet, according to {\bf Table 1}, by 
\be
\label{dela1035}
M_{rot}(\overline{35},J=3/2)-M_{rot}(\overline{10},J=1/2)={3\over 2\Theta_K}+{3\over 2\Theta_\pi}
\ee
which is about $750\,-\,800\,MeV$, i.e. considerably greater than quark model estimates \cite{clo}.
Some contribution to the mass difference of such partners comes also from $FSB$ mass terms, but
the mass splitting between components of $\{\overline{35}\}$ corresponding to $\{\overline{10}\}$
is smaller than that of $\{\overline{10}\}$ itself, almost twice: the value of $sin^2 \nu$ increases 
from $5/8$ to $3/4$, as shown in Table 6. These states are septuquarks, at least.

\begin{center}
\begin{tabular}{|l|l|l|l|l|l|}
\hline
$Baryon\,|N(p,q),Y,I,J>$   &$<s_\nu^2>_0$ & A & B & C &      \\

\hline
$\Theta^*\,|\overline{35},2,0,3/2>$&$0.625$&2423&2535 & 2487 &  \\ 
$N^*|\overline{35},1,1/2,3/2>   $&$0.667$&2481&2586 & 2548 &  \\

$\Sigma^*\,|\overline{35},0,1,3/2>$&$0.708$&2527&2628 & 2596 &  \\
$\Xi^{*}|\overline{35},-1,3/2,3/2>$&$0.750$&2557  & 2658 & 2627 &  \\
\hline
\hline
$N^*\,|27,1,1/2,3/2>$         &$0.643$  &1739 &1782 & 1783 &  \\ 
$\Sigma^*\,|27,0,1,3/2>$      &$0.679$  &1847 &1871 & 1896 &  \\ 
$\Lambda^*\,|27,0,0,3/2>$     &$0.714$  &1829 &1861 & 1876 &  \\
$\Xi^{*} |27,-1,1/2,3/2>$  &$0.768$  &1917 &1937 & 1969 &  \\
\hline  
$\Delta^*|35,1,3/2,5/2>$   &$0.438$&2054  & 2122 & 2137 &  \\

$\Sigma^*\,|35,0,1,5/2>$     &$0.542$&2123  & 2181 & 2209 &  \\
$\Xi^{*}|35,-1,1/2,5/2>$ &$0.646$&2186  & 2235 & 2275 &  \\
$\Omega^{*}\,|35,-2, 0,5/2>$ &$0.750$&2244  & 2286 & 2336 &  \\
\hline
\end{tabular}
\end{center}

{\bf Table 6.} {\tenrm Values of masses of the partners of the antidecuplet with $J=3/2$ within 
$\{\overline{35}\}$-plet; partners of the lowest octet within the $\{27\}$-plet, $J=3/2$; and
partners of the lowest decuplet within the $\{35\}$-plet, $J=5/2$. Rigid rotator approximation
has been used here according to \cite{wk,ufk}.}\\
 
There are also partners of other lowest multiplets, e.g. the partners of baryons octet 
$J^P=1/2^+$ with $J^P=3/2^+$ exist, the lowest one is contained within $\{27\}$-plet with $J^P=3/2^+$.
 The difference of rotation energies equals to 
\be
M_{rot}(27,J=3/2)-M_{rot}(8,J=1/2)={1\over \Theta_K}+{3\over 2\Theta_\pi},
\ee
so, about $580\,-\,620\,MeV$. More accurate numbers are presented in Table 6.
The partners of lowest $J=3/2$ decuplet with $J=5/2$ sit within $\{35\}$-plet and have the energies
shown also in Table 6, last 4 lines.

Generally, within complicated $SU(3)$ multiplets, like $\{27\}$ and $\{35\}$-plets, there are also 
partners of iso-multiplets with same spin and different isospins. For example, within $\{27\}$-plet, 
for $Y=0$
there are partners with isospin $I=2,\;I=1$ and $I=0$, spin $J=1/2$ and $J=3/2$; for $Y=1$ there are
states with $I=3/2$ and $1/2$ (see {\bf Figure}). For fixed value of spin $J=I_R$
states with different values of isospin $I$ have same rotational energy, their mass difference is 
due to $FSB$ 
terms, only. Such partners can be obtained within other approaches, see e.g. \cite{io}.
The spectrum of baryon states, including their partners, is rich, and interesting problem is 
to find correspondence with the spectrum arising from the quark models.
\section{Multibaryons with exoticness}
Numerous applications of the chiral soliton models to the properties of baryons have been widely 
discussed, mostly accepted and also criticized in literature. Another branch of these applications 
are the properties of states with baryon 
number greater than $1$, nuclei and/or multibaryons, and this issue is much less known and accepted.
The possibility to describe real nuclei as quantized chiral solitons appeared after discovery of
classical chiral field configurations bound relative to the transition to states with smaller
baryon numbers (history of this discovery and references can be found, e.g. in \cite{hms,vkax,kz}). 
One of recent results is successful description of the mass splittings of nuclear isotopes
with different values of isospin, or so called "symmetry energy" of nuclei \cite{ksm}. Some 
variation of the only parameter of the model, Skyrme constant, allowed to provide good 
description of data for atomic numbers up to $\sim 30$ and to predict binding energies (b.e.) of some 
neutron rich nuclides \cite{ksm}, in general agreement with other, more traditional, approaches.
The binding energies of light hypernuclei also can be calculated in general agreement with data
\cite{vk1}. Therefore, one can conclude that the chiral soliton approach provides  results
which are, at least, in qualitative agreement with existing nuclear physics data.

To obtain baryon states with definite quantum numbers, the quantization of classical configurations
should be performed. There are several quantization prescriptions described in the literature
to get the states with flavor quantum numbers, strangeness, charm or beauty.
Besides "rigid rotator" quantization scheme originating from works \cite{anw,g}, described and 
used above, there is also bound state approach \cite{ck,kk} and its simplified and very transparent
version \cite{kk,westk} which is convenient for estimating the energies of states with lowest flavor or
anti-flavor quantum numbers. This scheme has been used recently to find the spectrum of lowest
states with positive strangeness (beauty) or negative charm \cite{ksh}.

Within this quantization scheme the energy of state consists of two contributions. One, most important,
is the flavor excitation energy which is of the order of $N_c^0\sim 1$. Second component is the
correction of the order $\sim 1/N_c$ depending on the isospin of the state (hyperfine splitting
correction). 
In the leading order in $N_c$ the hamiltonian of the system can be written as
\be \label{hdd}
H =M_{cl} + 4 \Theta_{F,B} \Pi^\dagger \Pi + 
\biggl( {3\over 2}\Gamma_{SB} + {N_c^2B^2\over 16\,\Theta_{F,B}} \biggr) D^\dagger D
\ee
where $\Gamma_{SB}$ is given above by Eq. (\ref{GSB}), the moment of inertia 
$\Theta_{F,B}=\Theta_K$ for $B=1$, 2-component amplitude $D$ is deviation of
starting $SU(2)$ soliton into "flavor direction" which is believed to be small. Indeed, it can be
easily obtained that \be
|D| \sim [24 \Theta_{F,B} \Gamma_{SB} +N_c^2B^2]^{- 1/4},
\ee
i.e. it decreases with increasing $FSB$ mass and/or number of colors $N_c$. $\Pi$ is momentum 
canonically conjugate to variable $D$.
The relation takes place, $D^\dagger D \simeq (1-D_{88})/3 \,=\, s_\nu^2/2$ which is fulfilled
with good accuracy when $\nu$ is small.

This hamiltonian can be diagonalized and written in terms of flavor or antiflavor numbers 
\cite{kk,westk},
\be\label{hab}
H = M_{cl} + a^\dagger a \omega_{F,B} + b^\dagger b \bar{\omega}_{F,B}
\ee
$a^\dagger,\;b^\dagger$ being flavor and antiflavor creation operators, and flavor (antiflavor) 
excitation energies 
\be
\omega_{F,B}= {N_cB\over 8\Theta_{F,B}} \bigl(\mu_{F,B}-1\bigr), \qquad
\bar{\omega}_{F,B}={N_cB\over 8\Theta_{F,B}} \bigl(\mu_{F,B}+1\bigr),
\ee
\be
\label{mu}
\mu_{F,B}=\biggl[1 + {24\Theta_{F,B}\Gamma_{SB} \over (N_cB)^2}\biggr]^{1/2}.
\ee
At large $N_c$ the quantities $\mu_{F,B},\;\omega_{F,B}$ and $\bar{\omega}_{F,B}$ scale like 
$N_c^0\sim 1$. 
When $FSB$ is small, the expansion of $\mu_{F,B}$ can be made, and the flavor excitation energy 
\be \label{smallm}
\omega_{F,B}\simeq {3\Theta_{F,B}\Gamma_{SB}\over 2N_cB},
\ee
quadratically depends on the $FSB$ mass, because $\Gamma_{SB}\sim m_K^2$.
Further details and formulas can be found in \cite{westk,ksh}. 
\begin{center}
\begin{tabular}{|l|l|l|l|l|l|l|}
\hline
B & $\bar{\omega}_s$ & $\bar{\omega}_c $& $\bar{\omega}_b $& $\Delta\epsilon_s$& $\Delta\epsilon_c$&
 $\Delta\epsilon_b$\\
 \hline
1 &$591$ &$1750$ &$4940$ &---  &--- &---  \\
2 &$571$ &$1720$ &$4900$ &$ 45$ & $60$ & $90$\\
3 &$564$ &$1710$ &$4890$ &$ 65$ & $40$ & $50$ \\
4 &$567$ &$1710$ &$4870$ &$ 20$ & $15$ & $50$ \\
6 &$555$ &$1710$ &$4880$ &$ 55$ & $30$ & $40$ \\ 
8 &$553$ &$1710$ &$4890$ &$ 70$ & $30$ & $40$ \\
12&$547$ &$1720$ &$4910$ &$ 85$ & $30$ & $30$ \\
16&$541$ &$1720$ &$4930$ &$ 95$ & $30$ & $10 $ \\
20&$538$ &$1730$ &$4940$ &$100$ & $20$ & $-10$ \\
24&$536$ &$1730$ &$4960$ &$105$ & $20$ & $-20$ \\
\hline
\end{tabular}
\end{center}
{\bf Table 7.} The anti-flavor excitation energies (in $MeV$) and binding energies changes of
anti-flavored hypernuclei in comparison with ground states of ordinary nuclei (also in $MeV$) for 
baryon numbers up to $24$. \\

The energies of anti-flavors excitation and binding energies changes of flavored multibaryons
in comparison with ordinary nuclei presented in {\bf Table 7}, are taken from \cite{ksh}.
For anti-charm and anti-beauty the antiflavor excitation energies are considerably smaller
than masses of $D$ and $B$-mesons, correspondingly. It means that anti-charmed (anti-beautiful)
pentaquarks - their ground states - are bound relative to strong decays, since $1/N_c$ 
corrections are small for large mass of flavor (the masses of ground state pentaquarks 
are $\sim 2700\,MeV$ for anti-charm and $5880\,MeV $ for $\Theta_b$ \cite{ksh}). The property 
of binding of anti-charmed (-beautiful) pentaquarks is known really long ago \cite{rscoc,ohpm}.
The anti-charmed pentaquark observed recently with the mass $3099\,MeV$ \cite{akt} can be
some excitation of the ground state we discuss here, see, however \cite{karlip} where the
mass of the $\Theta_c$ was predicted to be $2985 \pm 50\, MeV$ within correlated quark model.

Some decrease of b.e. for anti-charm and anti-beauty, presented in Table 7, can be an artifact 
of approximations used to calculate them (rational map approximation \cite{hms}).
For baryon numbers $B\geq 10$
the rescaled (or nuclear) variant of the model can be used, which leads to increase of b.e. in 
comparison with nucleon variant of the model, by several tens of $MeV$ \cite{ksh}. It should be 
kept in mind 
that the mass of $\Theta^+$ hyperon within this particular variant of the model equals to $1588\,MeV$.
The accuracy of calculation is not better than $\sim 30-50 \,MeV$,  but deeply bound 
$\Theta$-hypernuclei should be expected for atomic numbers greater than $\sim 20$.
Similar results have been obtained also within more traditional potential or mean-field approaches
\cite{cabrera,zhong,vicen}, discussion of this issue and references can be found in \cite{lan}.

The increase of energy of exotic states in comparison with nonexotic ones was obtained for arbitrary
$N_c$ within rigid rotator model as well (Appendix of \cite{ksh}). It was found 
\be\label{odd}
\Delta E_{rot} = {N_cB+3\over 4\Theta_{F,B}}
\ee
for odd $B$-numbers, and for $B=1$ this coincides with above expression, second of (\ref{d108}).
For even $B$-numbers
\be\label{even}
\Delta E_{rot} = {N_cB+2\over 4\Theta_{F,B}}.
\ee
In derivation of these expressions it was assumed that ground states of nuclei and lowest states
of flavored multibaryons belong to $SU(3)$ multiplets $(p,q)$ with the lowest possible values of
$p$ \cite{ksh}, i.e. they have lowest allowed value of isospin - in general agreement with data. 
What is remarkable, the leading in $N_c$ contribution is the same as in
rigid oscillator model \cite{westk} used to make calculations in \cite{ksh}, where the difference
of anti-flavor and flavor excitation energies
\be 
\bar{\omega}_{F,B} - \omega_{F,B} = {N_cB\over 4\Theta_{F,B}}.
\ee
Evidently,  convergence of both quantization 
methods improves not only with increasing $N_c$, but also with increasing baryon number.

It should be noted that two different methods of quantization used in present paper, the rigid
(or soft) rotator used in previous sections, and rigid oscillator method, a variant of the 
bound state 
approach \cite{kk,westk}, are not identical and lead to different results for $N_c=3$. 
According to (\ref{smallm}), the mass 
splitting of decuplet is $\Delta_M(\{10\},RO) = 3\Gamma_{SB}/2$, whereas for rigid rotator it is
8 times smaller, according to previous results, see (\ref{HSB}), Table 2 and also Appendix below. 
For the octet of baryons the RO result for total
splitting is 4 times greater than the RR result. The RO method works well for exotic baryon and 
multibaryon states, but meets difficulties in describing the nonexotic components of $SU(3)$
multiplets which contain exotic states.

Another issue of interest could be the properties of classical chiral field configurations at 
large baryon numbers,
as obtained within the Skyrme model. Analytical evaluations performed in \cite{vk2} have shown that
these properties are quite universal: at large B-numbers multibaryons described within the Skyrme
model are spherical bubbles with the mass and B-number concentrated in their shell. The thickness
of this shell is approximately constant, about $t \sim 3.6/(F_\pi e)$, same is the average energy 
density in the shell, $\rho_{shell}\sim F_\pi^4 e^2$, if the mass term in the lagrangian is small 
enough. Both $t$ and $\rho$ do not depend on baryon number.
So, the bags of matter appear in this model, the properties of "material" which these bags are 
made of, follow from effective chiral lagrangian \cite{vk2} (in difference from traditional bag 
models where these properties are postulated, or introduced from phenomenological grounds). 
Although multibaryon configuations obtained in this way differ
from ordinary nuclei, by the form of their density distribution first of all, further modifications
of the model are possible, including modifications of the mass term \cite{marl2,vk2}, but I will 
not go into further details here.
\section{Conclusions and prospects}
The contradictive situation with observation of pentaquark states will be resolved, probably, 
within next few years.
Even if not all reported pentaquark states are confirmed, one could state that interesting 
branch of 
baryon spectroscopy appeared which will enlarge
 our knowledge about hadron structure.
If none of announced observed pentaquarks is confirmed, there remains still certain theoretical 
interest in understanding the structure of pentaquarks and correspondence between chiral soliton
and quark model descriptions.
Such states can appear as broader resonances at higher energies, as it was discussed previously 
\cite{j,hog,str,roi}.
Present discussion, certainly, puts more questions than gives answers.
The following problems and questions can be pointed out, many of them have been, of course, noted
in previous discussions \cite{hic2,azim}:

$*$ High statistics confirmation of existence of narrow pentaquarks seems to be necessary, 
especially for the resonances 
 $\Phi/\Xi_{3/2}$ and $\Theta_c$, see \cite{pr}.
 Some information about experiments performed 
or to be performed is contained in \cite{pr,hicks,hic2,kab2}.

$*$ Width determination is of great importance, $\Gamma \sim 1 \,MeV$ is not excluded and suggested
by analyses of scattering data,
 but would be difficult to explain it by theory: a special reason 
is necessary then.

$*$ Several missing components of multiplets remain to be found, for example, of considerable 
interest are:\\
in $\{\overline{10}\}$-plet: $\Phi/\Xi_{3/2}^+ \to \Xi^0 \;\pi^+,\;\; \Sigma^+\,\bar{K}^0$;\\
in $\{27\}$-plet: $\Theta^*_1\to NK$; $\Xi^*_{3/2} \to \Xi\pi$; $\Sigma_2 \to \Sigma\,\pi;\;$; 
$\Omega_1 \to \Omega\,\pi,\, \Xi \bar{K},$;\\
in $\{35\}$-plet: $\Omega_1^* \to \Omega\,\pi,\; \Xi \bar{K},\;  \Xi^* \bar{K}$; 
$\;\Delta_{5/2}\to N\pi\pi;\;$
$\Gamma_{S=-4} \to \Omega \bar{K}$, etc.

In the latter case the complication is due to the fact that most of interesting components of
$\{35\}$-plet are not available in octet-octet meson-baryon interaction.

$*$ Studies of cryptoexotics $(N^*,\; \Delta^*,\; \Xi^*...)$ are of interest as well, to complete
the picture of pentaquarks, more detailed discussion can be found in \cite{azim,araz}.

$*$ Spin and parity are {\it crucial} for cheking the validity of chiral soliton models
predictions. Negative parity of these states would provide big difficulties for their
interpretation as quantized topological solitons, although in any model it seems unrealistic
to get a narrow resonance, with $\Gamma \leq 10\,MeV$, decaying into S-wave state of meson and baryon
with energy release about $100\,MeV$.

$*$ As a result, better understanding of the structure of baryons and their wave functions will be 
reached.
The understanding of the possible important role of correlated diquarks and triquarks in the baryons 
wave functions could be the first example \cite{jw,dor}. The link of the soliton approach and quark models leads to
the conclusion that the effective masses of strange quark and antiquark within baryon states should
be considerably different and depend on the particular $SU(3)$ multiplet, and this can be another 
example. The difference of masses of strange antiquarks within $27$-plet and $35$-plet is so large,
that it looks as paradox.

$*$ As it was noted in literature \cite{coh,ikor}, predictions of chiral soliton models are not 
completely selfconsistent from the point of view of the $1/N_c$ expansion. 
In addition to problems considered in  \cite{coh,ikor}, we note, e.g., that 
the mass splitting between  octet and antidecuplet of baryons is of the order of $N_c^0\sim 1$, 
whereas the total mass splitting within octet or antidecuplet is of the order of $N_c$, as the 
classical soliton mass itself. 
There is also some inconsistency between rigid rotator and bound state quantization models, in 
particular the mass splittings within $SU(3)$ multiplets given by these models, differ 
considerably. These mass splittings
coincide at large $N_c$, in the leading in $1/N_c$ approximation, but for $N_c=3$ the bound state
approach in its present form (the rigid oscillator model, in particular \cite{westk}) 
gives much greater splittings when FSB mass is not large (see Appendix). In view of these difficulties,
the results obtained in the large $N_c$ limit, including e.g. some objections against chiral
soliton model results \cite{coh,ikor}, should be interpreted with great care and may not be valid for
the real $N_c=3$ world. 
Predictions of chiral soliton approach should be considered as a reasonable extrapolation, 
when one of states of interest is fitted. Results of this extrapolation are impressive sometime.

$*$ Other predictions of CSM are of interest besides those discussed in present paper, e.g.
supernarrow radiatively decaying dibaryon (JINR and INR experiments \cite{khr,fil}, see,
however, \cite{tam} where negative result was obtained for low values of dibaryon masses).

$*$ Chiral soliton models are good example of the field theoretical models which allow to obtain
the results of practical interest. They provide a possibility to describe not only baryons and
baryonic resonances, but also systems with large baryon numbers as "bags" of certain type, 
the properties of these bags are deduced from initial lagrangian \cite{vk2}.
 
Some important and interesting issues have not been considered here in view of restricted size of the
paper: exotic baryons production mechanisms and their properties; determination of spin-parity
and electromagnetic properties of these baryons, etc. Discussion and references can be found in
\cite{jm,azim}.

{\it Acknowledgments.} Computer programs for configuration mixing arranged by Bernd Schwesinger and
Herbert Weigel
(soft rotator approximation) and by Hans Walliser (rigid rotator) have been used in present paper.
I'm greatly indebted to Hans Walliser also for his help in numerical calculations, useful remarks
and criticism, to Tom Cohen and Igor Klebanov for helpful discussion of large $N_c$ subtleties,
to Igor Strakovsky for reading the manuscript and useful comments on experimental data
available now, and to Andrei Shunderuk for checking of some formulas and numerical results. E-mail 
conversations with Ya.I.Azimov, D.I.Diakonov, G.Holzwarth, R.L.Jaffe, S.Kabana, J.Trampetic, 
H.Weigel are acknowledged, as well as discussions with A.A.Andrianov, V.A.Andrianov, L.N.Bogdanova, 
K.G.Boreskov, B.L.Ioffe, B.Z.Kopeliovich, L.N.Lipatov, V.M.Lobashev, V.A.Matveev, Yu.V.Novozhilov, 
L.B. Okun', V.A.Rubakov, Yu.A.Simonov, K.A. Ter-Martirosyan.\\

\section{Appendix. Some properties of the large $N_c$ baryons.}
As it was discussed above, at large arbitrary (odd) number of colors $N_c$ baryon consists of
$N_c$ quarks in color singlet state, and there are totally $(N_c+1)/2$ nonexotic $SU(3)$ 
multiplets of baryons, from $[p,q]=[1,(N_c-1)/2]$, to $[p,q]=[N_c,0]$.
The hypercharge for arbitrary $N_c$ is $Y=N_cB/3 +S$ (\cite{g}, see also \cite{ikor,coh}). 
One possibility
for the choice of charges of quarks is as usual, $Q_u=2/3$, $Q_d=Q_s=-1/3$, see \cite{coh} e.g. 
In this case the electric 
charge defined by relation $Q=I_3 +Y/2$ is integer only if $N_c$ is multiple of $3$. Another 
possibility for electric charges was discussed in \cite{abb}.
where the supercharged quarks
and $SU(3)$ multiplets were considered. In this case $Q=I_3+Y/2 + B(3-N_c)/6$, the charges of quarks
follow from this expression at $B=1/N_c$:
\be
Q_u={1\over 2} + {1\over 2N_c},\qquad  Q_d=Q_s=-{1\over 2} + {1\over 2N_c}, 
\ee
and average charge of each baryonic $SU(3)$ multiplet, or supercharge,  equals to $\bar{Q}=(3-N_c)B/6$.

Below strangeness contents of baryons at large number of colors are presented, which define the
mass splittings of baryon multiplets within rigid rotator approximation.
For the multiplet $[p,q]=[1,(N_c-1)/2]$ which is analogue of the $N_c=3$ octet we obtained:
\bea \label{scoc}
<s_\nu^2>_{"N"} &=& {4(N_c+4)\over (N_c+3)(N_c+7)}, \qquad <s_\nu^2>_{"\Lambda"} = {6\over (N_c+7)},
\\ \nonumber
<s_\nu^2>_{"\Sigma"} &=& {2(3N_c+13)\over (N_c+3)(N_c+7)}, \qquad 
<s_\nu^2>_{"\Xi"} = {8\over (N_c+7)}.
\eea
Evidently, the Gell-Mann---Okubo relation $M_\Sigma + 3M_\Lambda =2(M_N+M_\Xi)$ is fulfilled
for these values of $<s_\nu^2>$. The increase of $<s_\nu^2>$ per unit of strangeness within "octet"
of baryons equals, in average, at large $N_c$
\be
\Delta(<s_\nu^2>,\delta |S|=1,"8") \simeq {2\over N_c}\left (1-{8\over N_c} \right ).
\ee
For the components of "decuplet" $[p,q]=[3,(N_c-3)/2]$ we obtain
\bea \label{scdec}
<s_\nu^2>_{"\Delta"} &=&  {4(N_c+4)\over (N_c+1)(N_c+9)}, \qquad 
<s_\nu^2>_{"\Sigma*"} =  {2(3N_c+7)\over (N_c+1)(N_c+9)}, \\ \nonumber
<s_\nu^2>_{"\Xi*"}  &=&  {4(2N_c+3)\over (N_c+1)(N_c+9)}, \qquad 
<s_\nu^2>_{"\Omega"} = {10\over (N_c+9)},  .
\eea
Equidistant behaviour of "decuplet" components can be noted, with a step 
\bea
\Delta(<s_\nu^2>,\delta |S|=1,"10")& = &2(N_c-1)/[(N_c+1)(N_c+9)] \\ \nonumber
    & \simeq & {2\over N_c}\left (1-{11\over N_c}+{101\over N_c^2}\right ).
\eea
At large $N_c$ average splittings within "octet" and "decuplet" coincide and equal to $2/N_c$, 
but preasymptotic corrections $\sim 1/N_c^2$ are different, making splitting within "decuplet"
smaller than within "octet", in qualitative agreement with observations for octet and 
decuplet in real world.

For the $\Theta^+ \in "\{\overline{10}\}"$ it is easy to obtain
\be
<s_\nu^2>_{"\Theta"} = {6\over (N_c+9)},
\ee
and for the $\Theta_1 \in "\{27\}"$
\be
<s_\nu^2>_{"\Theta_1"} = {2(3N+23)\over (N_c+5)(N_c+11)},
\ee
which is slifgtly greater than for $\Theta$.

At large $N_c$ it is a matter of simple algebra to establish that there is equidistant
behaviour of strangeness contents (recall that $<s_\nu^2>_B = 2 SC_B$) for the components of 
non-exotic multiplets, with not large (fixed) values of strangeness $S$:
\be
   <s_\nu^2>_{(p\sim 1,q\sim N_c/2, |S|\sim 1)} \simeq 2 {2+|S| \over N_c}.
\ee
As a result, in the limit $N_c \to \infty $ the mass splittings between adjacent components of 
"octet", "decuplet" and other
nearest multiplets coincide with those obtained within rigid oscillator approximation \cite{westk}.
However, $1/N_c^2$ corrections to these asymptotic values of mass splittings are large, the expansion 
parameter is about $\sim 11/N_c$ or $\sim 8/N_c$. These corrections lead to the decrease of the 
mass splittings
between adjacent components of these multiplets, and this effect becomes of the order 1 when
$|S| \sim N_c$. As a result, the total mass splittings of the whole multiplets are smaller than within
RO approximation by numerical factors. For the real world, $N_c=3$, the mass splitting of decuplet
is 8 times smaller for rigid rotor approximation than for rigid oscillator (when FSB mass is small
and hyperfine splitting correction of the order $\sim 1/N_c$ is not included),
and 4 times smaller for the octet. Hyperfine splitting correction as found in \cite{kk,westk} 
decreases the mass splitting within RO model, but not sufficient. It is possible to modify
the next to leading in $1/N_c$ contributions to the mass splittings by means of appropriate
resolution of the operator ordering ambiguity within RO and to remove the difference from
RR model  \cite{kleko}. However, it is not clear how to make extrapolation to realistic value
$N_c=3$, and what is the influence of such modification on results and conclusions of paper
\cite{ikor}. This illustrates well that 
although both methods, RR and RO, converge at large $N_c$, small $m_K$ and fixed not large values of 
strangeness, in real world there is considerable difference between both approaches.Recently the 
paper \cite{chl} echoed the difficulties of extrapolation of results obtained in large $N_c$ world 
to the realistic $N_c=3$ world, at least for some physical observables

When $m_K$ is large (as for charm or beauty quantum numbers) the flavor excitation energies for
RO method depend linearly on $m_D$, which looks much more realistic than for RR method, and the
bound state approach \cite{ck,westk} is more preferable.\\
\\
NOTES ADDED in PROOF. The higher statistics study of the positively charged kaon interactions
in the Xe bubble chamber \cite{VVB} reinforced the evidence of the DIANA Collaboration for
the production of the $\Theta^+$ hyperon, with the confidence level from 4.3 to 7.3
standard deviations, depending on the estimation method. At the same time, the CLAS Collaboration 
in recent high statistics experiment \cite{BMcK} disavowed their previous result on the observation
of $\Theta^+$ in the photoproduction reaction on deuterons.

Detailed calculations of the strangeness contents of all components of exotic multiplets of
baryons (pentaquarks) have been performed recently at an arbitrary number of colors $N_c$
\cite{VBKAMS} within the rigid rotator model. The leading terms in the $1/N_c$ expansion
for the positive strangeness states coincide with those in the rigid oscillator model,
but the next-to-leading order terms differ essentially from those obtained within the
RO model in its commonly accepted version, as for the "octet" and "decuplet" of baryons
discussed in \cite{kleko} and in present paper.\\
\\ 
{\elevenbf References}


\begin{thebibliography}{111}
\bibitem{PDG} PDG Collab., Phys. Lett. B592 (2004); J. Phys. G 33, 1 (2006)

\vspace{-1.mm}
\bibitem{1} T. Nakano et al (LEPS Collab.), Phys. Rev. Lett. 91, 012002
(2003); hep-ex/0301020
\vspace{-1.mm}
\bibitem{2} V.V. Barmin et al (DIANA Collab.) Yad. Fiz. 66, 1763 (2003); hep-ex/0304040
\vspace{-1.mm}
\bibitem{3} S. Stepanyan et al (CLAS Collab.) Phys. Rev. Lett. 91, 252001 (2003)
\vspace{-1.mm}
\bibitem{4} J. Barth et al (SAPHIR Collab.) hep-ex/0307083
; Phys. Lett. B572, 127 (2003)
\vspace{-1.mm}
\bibitem{5} A.E. Asratyan, A.G. Dolgolenko and M.A. Kubantsev, Yad.Fiz. 67, 704 (2004)
\vspace{-1.mm}
\bibitem{6} V. Kubarovsky et al (CLAS Collab.) Phys. Rev. Lett. 92, 032001
 (2004)
\vspace{-1.mm}
\bibitem{7} S. Chekanov et al (ZEUS Collab.) Phys. Lett. B591, 7 (2004)
\vspace{-1.mm}
\bibitem{8} A. Airapetian et al (HERMES Collab.) Phys. Lett. B585, 213 (2004)
\vspace{-1.mm}
\bibitem{9} A. Aleev et al (SVD Collab.) hep-ex/0401024
, Phys. Atom. Nucl. 68, 974 (2005)
\vspace{-1.mm}
\bibitem{10} M. Abdel-Bary et al (COSY-TOF Collab.) hep-ex/0403011
; Phys. Lett. B595, 127 (2004)
\vspace{-1.mm}
\bibitem{troyan} Yu.A. Troyan et al (JINR $H_2$ bubble chamber Collab.), hep-ex/0404003
\vspace{-1.mm}
\bibitem{bdv} M. Battaglieri, R. De Vita, talk at the meeting NSTAR-2004 
\vspace{-1.mm}
\bibitem{asw} R. Arndt, I. Strakovsky and R. Workman, Phys. Rev. C68, 042201 (2003);
R. Workman, R. Arndt, I. Strakovsky, D. Manley and J. Tulpan, Phys. Rev. C70, 028201 (2004)
\vspace{-1.mm}
\bibitem{gib} R.N. Cahn and G.H. Trilling, Phys. Rev. D69, 011501 (2004)
\vspace{-1.mm}
\bibitem{poc} J. Pochodzalla, {\it Pentaquarks: facts and mysteries, or Sisyphus at work.} 
hep-ex/0406077
\vspace{-1.mm}
\bibitem{hicks} K. Hicks, {\it Workshop Summary: Experiment}, hep-ex/0501018
\vspace{-1.mm}
\bibitem{azim} Ya.I. Azimov, R.A. Arndt, I.I. Strakovsky, R.L. Workman and K. Goeke,
Eur. Phys. J. A26, 79 (2005)
\vspace{-1.mm}
\bibitem{hic2} K.H. Hicks, {\it Experimental search for pentaquarks},
Prog. Part. Nucl. Phys. 55, 647 (2005)
\vspace{-1.mm}
\bibitem{devita} M. Battaglieri, R. De Vita, V. Kubarovsky, D. Weygand and the CLAS Collab.,
Talk at the APS Meeting, April 16  (2005)
\vspace{-1.mm}
\bibitem{alt} C. Alt et al (NA49 Collab.) Phys. Rev. Lett. 92, 042003 (2004); hep-ex/0310014
\vspace{-1.mm}
\bibitem{comp} E.S. Ageev et al (COMPASS Collab.) hep-ex/0503033, 
Eur. Phys. J. C41, 469 (2005)
\vspace{-1.mm}
\bibitem{spen} J. Spengler (for HERA-B Collab.) hep-ex/0504038, Acta Phys. Polon.
B36, 2223 (2005)
\vspace{-1.mm}
\bibitem{akt} A. Aktas et al (H1 Collab.) Phys. Lett. B588, 17 (2004); hep-ex/0403017
;
C. Risler (for H1 Collab.), hep-ex/0506077
\vspace{-1.mm}
\bibitem{zeus} U. Karshon (for ZEUS Collab.) hep-ex/0407004; hep-ex/0410029; 
S. Chekanov et al (ZEUS Collab.)
 hep-ex/0409033; Eur. Phys. J. C38, 29 (2004)
\vspace{-1.5mm}
\bibitem{rip} M. Ripani et al (CLAS Collab.), Phys. Rev. Lett. 91, 022002 (2003);
 
hep-ex/0304034
\vspace{-1.mm}
\bibitem{kab} S. Kabana (for STAR Collab.) hep-ex/0406032
\vspace{-1.mm}
\bibitem{aer} P. Aslanyan, V. Emelyanenko and G. Rikhvitskaya, hep-ex/0504026
\vspace{-1.mm}
\bibitem{pr} V. Kubarovsky,  P. Stoler (for CLAS Collab.) hep-ex/0409025; 
P. Rossi, Nucl. Phys. A752, 111 (2005)
\vspace{-1.mm}
\bibitem{kab2} S. Kabana, AIP Conf. Proc. 739, 181 (2005); ibid. 756, 195 (2005)
\vspace{-1.mm}
\bibitem{j} R.L. Jaffe, SLAC-PUB-1774
; Phys. Rev. D15, 267, 281 (1977)
\vspace{-1.mm}
\bibitem{hog} H. Hogaasen and P. Sorba, Nucl. Phys. B145, 119 (1978)
\vspace{-1.mm}
\bibitem{str} D. Strottman, Phys.Rev. D20, 748 (1979)
\vspace{-1.mm}
\bibitem{roi} C. Roiesnel, Phys. Rev. D20, 1646 (1979)
\vspace{-1.mm}
\bibitem{mc} A. Manohar, Nucl. Phys. B248, 19 (1984); M. Chemtob, Nucl.Phys. B256, 600 (1985)
\vspace{-1.mm}
\bibitem{bieden} L.S. Biedenharn and Y. Dothan, {\it From SU(3) to Gravity}
(Ne'eman Festschrift), Cambridge University Press, Cambridge (1986)
\vspace{-1.mm}
\bibitem{km} M. Karliner and M.P. Mattis, Phys. Rev. D34, 1991 (1986)
\vspace{-1.mm}
\bibitem{mic} M. Praszalowicz, Proc. of the Workshop {\it Skyrmions and Anomalies}
Krakow, Poland, 20-24 Febr. 1987, World Scientific, Ed. M.Jezabek, 
M.Praszalowicz, p.112.
\vspace{-1.mm}
\bibitem{g} E. Guadagnini, Nucl.Phys. B236, 35 (1984)
\vspace{-1.mm}
\bibitem{vk} V. Kopeliovich, {\it On exotic systems of baryons in chiral soliton models.}
NORDITA Preprint 90/55 NP (1990); Phys. Lett. B259, 234 (1991)
\vspace{-1.mm}
\bibitem{jenm} E. Jenkins and A.V. Manohar, hep-ph/0401190, Phys. Rev. Lett. 93, 022001 (2004);
hep-ph/0402150, Phys. Rev. D70, 034023 (2004)
\vspace{-1.mm}
\bibitem{hw} H. Walliser, {\it An Extension of the Standard Skyrme model}, Proc. of the Workshop 
{\it Baryon as Skyrme Soliton},
 Siegen, Germany, 28-30 Sept. 1992, World Scientific, 
Ed. G.Holzwarth, p.247;
 Nucl. Phys. A548, 649 (1992)
\vspace{-1.mm}
\bibitem{dpp} D. Diakonov, V. Petrov and M. Polyakov, Z. Phys. A359, 305 (1997)
\vspace{-1.mm}
\bibitem{wei} H. Weigel, Eur. Phys. J. A2, 391 (1998)
\vspace{-1.mm}
\bibitem{wk} H. Walliser and V.B. Kopeliovich, JETP 97, 433 (2003)
; hep-ph/0304058
\vspace{-1.mm}
\bibitem{mic2} M. Praszalowicz, Phys. Lett.B575, 234 (2003); hep-ph/0308114
\vspace{-1.mm}
\bibitem{bfk} D. Borisyuk, M. Faber and A. Kobushkin, hep-ph/0307370; Ukr. J. Phys. 49, 
944 (2004)
\vspace{-1.mm}
\bibitem{mic3} M. Praszalowicz, Phys. Lett. B583, 96 (2004); hep-ph/0311230;
Acta Phys. Polon. B35, 1625 (2004); hep-ph/0402038
\vspace{-1.mm}
\bibitem{wuma} B. Wu and B.Q. Ma, hep-ph/0312041; hep-ph/0408121
\vspace{-1.mm}
\bibitem{ell} J. Ellis, M. Karliner and M. Praszalowicz, JHEP 0405002 (2004)
\vspace{-1.mm}
\bibitem{wei1} H. Weigel, Eur. Phys. J. A21, 133 (2004); hep-ph/0404173; hep-ph/0405285
\vspace{-1.mm}
\bibitem{tram} G. Duplancic, H. Pasagic and J. Trampetic, Phys. Rev. D70, 077504 (2004);
JHEP 0407, 027 (2004)
\vspace{-1.mm}
\bibitem{mix} H.K. Lee and H.Y. Park, hep-ph/0406051
\vspace{-1.mm}
\bibitem{kl} M. Karliner and H.J. Lipkin, hep-ph/0307243; Phys. Lett. B575, 249 (2003)
\vspace{-1.mm}
\bibitem{jw} R. Jaffe and F. Wilczek, Phys. Rev. Lett. 91, 232003 (2003); hep-ph/0307341
\vspace{-1.mm}
\bibitem{clo} F. Close, {\it The End of the Constituent Quark model?, Summary talk at the Baryon 03
Conf.}, AIP Conf. Proc. 717, 919 (2004), hep-ph/0311087
\vspace{-1.mm}
\bibitem{clodu} J.J. Dudek and F.E. Close, Phys. Lett. B583, 278 (2004), hep-ph/0311258; 
F.E. Close and J.J. Dudek, Phys. Lett. B586, 75 (2004), hep-ph/0401192
\vspace{-1mm}
\bibitem{wil} F. Wilczek, {\it Diquarks as Inspiration and as Objects}, hep-ph/0409168
\vspace{-1.mm}
\bibitem{clo2} F. Close, {\it Hadron Spectroscopy (theory): Diquarks, Tetraquarks, Pentaquarks and
no quarks, Plenary talk at 32nd International Conf. on High Energy Physics, ICHEP 04, Bejing, China,
August 2004}, hep-ph/0411396
\vspace{-1.mm}
\bibitem{io} S.-L. Zhu, Phys. Rev. Lett. 91,
 232002 (2003);
 B.L. Ioffe and A.G. Oganesian, 
Pis'ma v ZhETF 80, 439 (2004), hep-ph/0408152; R.D. Matheus and S. Narison, hep-ph/0412063;
\bibitem{dor} N.I. Kochelev, H.-J. Lee and V. Vento, Phys. Lett. B594, 87 (2004); 
A.E. Dorokhov and N.I. Kochelev, hep-ph/0411362; H-J. Lee, N.I. Kochelev and V. Vento, 
Phys. Lett. B610, 50 (2005), hep-ph/0412127; hep-ph/0506250
\vspace{-1.mm}
\bibitem{csi}  F. Csikor et al, hep-lat/0309090; JHEP 0311, 070 (2003); hep-lat/0503012
\vspace{-1.mm}
\bibitem{ssa} S. Sasaki, Phys. Rev. Lett. 93, 152001 (2004); hep-lat/0410016;
 T.-W. Chiu and T.-H. Hsieh, hep-ph/0403020
\vspace{-1.mm}
\bibitem{jm} B.K. Jennings, K. Maltman, Phys. Rev. D69, 094020 (2004), hep-ph/0308286
; 
 D. Diakonov,
 hep-ph/0406043; S.-L. Zhu, hep-ph/0406204; M. Oka, hep-ph/0406211;
M. Karliner, Acta Phys. Polon. B35, 3055 (2004); Int. J. Mod. Phys. A20, 199 (2005); 
R.L. Jaffe, Phys. Rep. 409, 1 (2005); M. Praszalowicz and K. Goeke, hep-ph/0506041;
A. Hosaka, hep-ph/0506138; I.M. Narodetsky, Yad. Fiz. 68, 780 (2005)
\vspace{-1.mm}
\bibitem{ufk} V. Kopeliovich, Physics Uspekhi, 47, 309 (2004)
; hep-ph/0310071
\vspace{-1.mm}
\bibitem{coh} T. Cohen, Phys. Lett. B581, 175 (2004), hep-ph/0309111; hep-ph/0312191
\vspace{-1.mm}
\bibitem{coh3} T.D. Cohen and R.F. Lebed, Phys. Lett. B578, 150 (2004), hep-ph/0309150
\vspace{-1.mm}
\bibitem{ikor} N. Itzhaki, I.R. Klebanov, P. Ouyang and L. Rastelli, Nucl. Phys. B684, 264 (2004);
hep-ph/0309305
\vspace{-1.mm}
\bibitem{sk} T.H.R. Skyrme, Proc. Roy. Soc. A260, 127 (1961)
\vspace{-1.mm}
\bibitem{wit} E. Witten, Nucl. Phys. B223, 433 (1983)
\vspace{-1.mm}
\bibitem{de} D.I. Diakonov and M. Eides, JETP Lett. 38, 433 (1983)
\vspace{-1.mm}
\bibitem{bal} J. Balog, Phys. Lett. B149, 197 (1984)
\vspace{-1.mm}
\bibitem{aa} A.A. Andrianov, V.A. Andrianov, Yu. V. Novozhilov and V.Yu. Novozhilov, Lett.
Math. Phys. 11, 217 (1986); Phys. Lett. B186, 401 (1987); ibid. B203, 160 (1988)
\vspace{-1.mm}
\bibitem{aj} A. Jackson et al, Phys. Lett. B154, 101 (1985)
\vspace{-1.mm}
\bibitem{marl} L. Marleau, Phys. Lett. B235, 141 (1990); 
Phys. Rev. D45, 1776 (1992); L. Marleau and J-F. Rivard, Phys. Rev. D63, 036007 (2001)
\vspace{-1.mm}
\bibitem{marl2} L. Marleau, Phys. Rev. D43, 885 (1991)
\vspace{-1.mm}
\bibitem{anw} G. Adkins, C. Nappi and E. Witten, Nucl. Phys. B228, 552 (1983)
\vspace{-1.mm}
\bibitem{vkax} V.B. Kopeliovich, Sov.J.Nucl.Phys. 47, 949 (1988); E. Braaten and L. Carson, 
Phys. Rev. D38, 3525 (1988)
\vspace{-1.mm}
\bibitem{dp} D. Diakonov, V. Petrov, Phys. Rev. D69, 094011 (2004)
\vspace{-1.mm}
\bibitem{pavar} M.M. Pavan, R.A. Arndt, I.I. Strakowsky and R.L. Workman, {\it Proc. of
 the 9th International Symposium on Meson-Nucleon Physics and Structure of Nucleon
 (MENU2001), Washington, DC, USA, July 26-31, 2001}, ed. by H.~Haberzettl and W.~Briscoe,
 $\pi$N~Newslett. 16, 110 (2002); hep-ph/0111066
\vspace{-1.mm}
\bibitem{kuz} V. Kuznetsov (for GRAAL Collab.), hep-ex/0409032
\vspace{-1.mm}
\bibitem{araz} R.A. Arndt, Ya.I. Azimov, M.V. Polyakov, I.I. Strakovsky and R.L. Workman, 
Phys. Rev. C69, 035208 (2004), nucl-th/0312126
\vspace{-1.mm}
\bibitem{arbri} R.A. Arndt et al, Phys. Rev. C69, 035213 (2004), nucl-th/0311089
\vspace{-1.mm}
\bibitem{schw} B. Schwesinger and H. Weigel, Phys. Lett. B267, 438 (1991)
\vspace{-1.mm}
\bibitem{mou} B. Moussalam, Ann. of Phys. (N.Y.) 225, 264 (1993); 
F. Meier, H. Walliser, Phys. Rept. 289, 383 (1997)

\vspace{-1.mm}
\bibitem{kss} V. Kopeliovich, B. Schwesinger and B. Stern, Nucl. Phys. A549, 485
 (1992)
\vspace{-1.mm}
\bibitem{abb} A. Abbas, Phys. Lett. B238, 344 (1990); Phys. Lett. B503, 81 (2001); H. Walliser,
unpublished.
\vspace{-1.mm}
\bibitem{oh} Y. Oh and H. Kim, Phys. Rev. D70, 094022, (2004), hep-ph/0405010; hep-ph/0409358
\vspace{-1.mm}
\bibitem{ck} C.G. Callan and I.R. Klebanov, Nucl. Phys. B262, 365 (1985);
N. Scoccola et al, Phys. Lett. B201, 425 (1988)

\vspace{-1.mm}
\bibitem{kk} D. Kaplan and I.R. Klebanov, Nucl. Phys. B335, 45 (1990)
\vspace{-1.mm}
\bibitem{westk} K.M. Westerberg and I.R. Klebanov, Phys. Rev. D50, 5834 (1994)
\vspace{-1.mm}
\bibitem{hms} C. Houghton, N. Manton and P. Sutcliffe, Nucl. Phys. B510, 507 (1998)
\vspace{-1.mm}
\bibitem{kz} V.B. Kopeliovich, W.J. Zakrzewski, JETP Lett, 69, 721 (1999);
Eur. Phys. J C18, 369 (2000); V.B. Kopeliovich, JETP, 93, 435 (2001)
\vspace{-1.mm}
\bibitem{vk2} V.B. Kopeliovich, J. Phys. G28, 103 (2002); JETP Lett. 73, 587 (2001); 
V. Kopeliovich, B. Piette and W. Zakrzewski, hep-th/0503127
\vspace{-1.mm}
\bibitem{ksm} V. Kopeliovich, A. Shunderuk and G. Matushko, nucl-th/0404020, Yad. Fiz. 
(2005) in press.
\vspace{-1.mm}
\bibitem{vk1} V.B. Kopeliovich, JETP 96, 782 (2003); nucl-th/0209040
\vspace{-1.mm}
\bibitem{ksh} V. Kopeliovich and A. Shunderuk,  JETP 100, 929 (2005)
, nucl-th/0409010
\vspace{-1.mm}
\bibitem{rscoc} D.O. Riska and N.N. Scoccola, Phys. Lett. B299, 338 (1993)
\vspace{-1.mm}
\bibitem{ohpm} Y. Oh, B.Y. Park and D.P. Min, Phys. Lett. B331, 362(1994);
Phys. Rev. D50, 3350 (1994)
\vspace{-1.mm}
\bibitem{karlip} M. Karliner and H.J. Lipkin, hep-ph/0307343
\vspace{-1.mm}
\bibitem{cabrera} D. Cabrera et al, Phys. Lett. B608, 231 (2005)
\vspace{-1.mm}
\bibitem{zhong} X. Zhong et al, Phys. Rev. C71, 015206 (2006)
\vspace{-1.mm}
\bibitem{vicen} M.J. Vicente Vacas et al, nucl-th/0410056
\vspace{-1.mm}
\bibitem{lan} D.E. Lanskoy, nucl-th/0411004
\vspace{-1.mm}
\bibitem{khr} A. Khrykin et al, Phys. Rev. C64, 034002 (2001)
\vspace{-1.mm}
\bibitem{fil} L.V. Filkov et al, Eur. Phys. J. A12, 369 (2001)
\vspace{-1.mm}

\bibitem{tam} A. Tamii et al, Phys. Rev. C65, 047001 (2002)
\vspace{-1.mm}
\bibitem{kleko} I.R. Klebanov and V.B. Kopeliovich, {\it "Large $N_c$ comparison of the rigid 
oscillator and rotator approaches to Skyrmions"}, June 2005, unpublished notes.
\vspace{-1.mm}
\bibitem{chl} T.D. Cohen, P.M. Hohler and R.F. Lebed, Phys. Rev. D72, 074010 (2005)
\bibitem{VVB} V.V. Barmin et al. (DIANA Collab.), hep-ex/0603017
\bibitem{BMcK} B. VcKinnon et al. (CLAS Collab.), Phys. Rev. Lett. 96, 212001 (2006)
\bibitem{VBKAMS} V.B. Kopeliovich and A.M. Shunderuk, Phys. Rev. D73, 094018 (2006)
\end{thebibliography}
\end{document}